\documentclass[manuscript]{aastex}

\usepackage{longtable}
\usepackage[usenames]{color}

\newcommand{\ltsimeq}{\raisebox{-0.6ex}{$\,\stackrel
        {\raisebox{-.2ex}{$\textstyle <$}}{\sim}\,$}}
\newcommand{\gtsimeq}{\raisebox{-0.6ex}{$\,\stackrel
        {\raisebox{-.2ex}{$\textstyle >$}}{\sim}\,$}}

\slugcomment{AJ revised ver \today}

\received{}
\revised{}
\accepted{}

\shorttitle{MIPSGAL and Taurus Asteroid Surveys}
\shortauthors{Ryan et al.}

\begin{document}
\title{THE KILOMETER-SIZED MAIN BELT ASTEROID POPULATION REVEALED BY SPITZER}


\author{ERIN LEE RYAN\altaffilmark{1}, DONALD R. MIZUNO\altaffilmark{2},
SACHINDEV S. SHENOY\altaffilmark{3}, \\ CHARLES E. WOODWARD\altaffilmark{4},
SEAN CAREY\altaffilmark{5}, ALBERTO NORIEGA-CRESPO\altaffilmark{5}, \\ 
KATHLEEN E. KRAEMER\altaffilmark{6}, STEPHAN D. PRICE\altaffilmark{6}}

\altaffiltext{1}{Minnesota Institute for Astrophysics, University of 
Minnesota, 116 Church Street, S.~E., Minneapolis, MN 55455, 
\textit{erinleeryan@gmail.com}}

\altaffiltext{2}{Institute for Scientific Research, Boston College,
140 Commonwealth Ave, Chestnut Hill, MA 02476-3862}

\altaffiltext{3}{ORAU -- NASA Ames Research Center, MS 245-6,
Moffett Field, CA, 94035-0001}

\altaffiltext{4}{Minnesota Institute for Astrophysics, University of 
Minnesota, 116 Church Street, S.~E., Minneapolis, MN 55455}

\altaffiltext{5}{Spitzer Science Center, MS 220-6, California Institute of
Technology, Pasadena, CA 91125}

\altaffiltext{6}{Institute for Scientific Research, Boston College,
855 Centre Street, Newton MA, 02459}

\begin{abstract}

Multi-epoch \textit{Spitzer Space Telescope} 24~\micron{} data 
is utilized from the MIPSGAL and Taurus Legacy surveys to detect 
asteroids based on their relative motion. These infrared detections 
are matched to known asteroids and rotationally averaged 
diameters and albedos are 
derived using the Near Earth Asteroid Model (NEATM) in conjunction with 
Monte Carlo simulations for 1835 asteroids ranging in size from 0.2 to 
143.6~km. A small subsample of these objects was also 
detected by IRAS or MSX and the single wavelength albedo and diameter
fits derived from this data are within 5\% of the IRAS and/or MSX 
derived albedos and diameters demonstrating the robustness of our 
technique. The mean geometric albedo of the small main belt asteroids 
in this sample is p$_{V} = 0.138$ with a sample standard 
deviation of 0.105.
The albedo distribution of this sample is far more diverse than the 
\textit{IRAS} or \textit{MSX} samples. The cumulative 
size-frequency distribution of 
asteroids in the main belt at small diameters is directly derived. Completeness limits of the optical and infrared 
surveys are discussed.

\end{abstract}


\section{INTRODUCTION\label{mipsgal_intro}}

Planetesimals are increasingly recognized as the 
evolutionary lynch pins for models of planet formation within the 
solar system. Their demographics, compositions, and dynamical 
attributes are imprints of our circumstellar ecosystem extant at the 
epoch of planet building that likely reflect the general conditions in 
exo-planetary disks. From the study of asteroids as relics of the 
early period of planet building, insight can be gained into 
the accretion processes and the initial composition of the 
proto-planetary disk. 

Previous asteroid surveys performed with \textit{IRAS} \citep{iras02} 
and \textit{MSX} \citep{msx02} enabled estimates of the albedo 
and diameter 
distributions of large main belt asteroids (MBAs). These surveys were flux 
limited to an asteroid diameter threshold of $\gtsimeq 10$~km; 
however, they still produced albedo and diameter estimates for $\simeq 
2000$ asteroids. Recently, the NEOWISE survey \cite{masiero2011} 
has released a preliminary catalog of albedos and diameters 
for $\simeq 10^{6}$ MBAs. This ensemble provided critical observational 
constraints for collisional models used to follow the evolution of 
planetesimals over the lifetime of the solar system 
\citep[e.g.,][]{bottke05}. These models use mean albedos for 
asteroids and their optical absolute magnitudes to generate the 
current day size-frequency distribution of 
asteroids. However, uncertainty exists regarding whether or not 
there is a tight and narrowly defined correlation between the 
albedos and diameters of asteroids. For instance, the \textit{IRAS} 
survey suggests that the range of asteroid albedos becomes more 
diverse with decreasing diameter. 

Compositional studies of main belt asteroids are utilized to explore 
whether or not our proto-planetary disk was contaminated by supernova 
products such as $^{26}$Al. Early compositional studies 
\citep{gradie1982} suggested evidence of a compositional gradient 
as a function of semi-major axis in the main belt - from highly 
thermally altered compositions in the inner belt to non-thermally 
altered compositions in the outer main belt. This gradient was 
attributed to parent body melting due to heating by the decay of 
radioactive isotopes \citep{grimmmcsween93,McSween2002}, 
and many models invoking this mechanism produced significant numbers 
of small thermally unaltered bodies in the inner main belt 
with diameters less than 20 km. However, 
this population has yet to be observed. For instance, the IRAS 
survey is incomplete for asteroids $< 20$~km at any zone in the main
belt.  Possibly, these small bodies 
were destroyed via mutual collisions \citep{Davis1989}, yet recent 
analysis of the Sloan Moving Object Catalog 
\citep[SMOC;][]{carvano2010} indicates that many small 
($\ltsimeq 10$~km) dark asteroids were missed in prior asteroid 
surveys. In addition the SMOC data indicate that colors of small 
main belt asteroids display 
significant compositional diversity as a function of semi-major 
axis, rather than the zoning present in the large asteroid population. 

The unique and unparalleled $\mu$Jy point-source flux density 
sensitivity of the \textit{Spitzer Space Telescope} during the 
cryogenic mission has enabled detection of faint asteroids with 
diameters as small as $\simeq 1$~km at high signal to noise in both
targeted surveys and serendipitous fields along the ecliptic. Here we 
utilize data from the MIPSGAL Galactic Plane survey and the Taurus 
Molecular Cloud survey to 
investigate the albedo behaviour of small asteroids, with a specific 
objective to determine whether or not the small ($\simeq 1$~km) small 
main belt asteroids have the same mean albedo and spatial albedo 
distribution as the large ($\ge 10$~km) main belt asteroids 
populations detected in earlier \textit{IRAS} and \textit{MSX} 
surveys. We use derived diameters from our MIPSGAL and Taurus 
catalogs to establish the size-frequency 
distribution of small main belt asteroids, and to 
assess whether the size-frequency distribution is functionally 
dependent on the heliocentric distance and/or composition. 

In section \ref{sec:observ_mips}, we briefly describe the mid-infrared 
(IR) surveys that were data-mined from the \textit{Spitzer} archive 
to produce our asteroid catalog. Section~\ref{sec:alb_det} discusses 
our approach to deriving asteroid albedos and diameters, while 
Section~\ref{sec:modeling_results} discusses our thermal modeling 
results, survey completeness limits, comparisons to prior 
\textit{IRAS} albedo catalogs of MBAs, as well as an examination 
of main belt albedo gradients, dynamical family albedos within the 
main belt and the overall bulk size-frequency distribution of 
asteroids. We conclude in Section~\ref{sec:conclusions}.

\section{MIPSGAL AND TAURUS SURVEYS \label{sec:observ_mips}}

The two \textit{Spitzer} surveys studied in this paper were 
selected via three criteria: multi-epoch 24~\micron{} data taken with 
epoch separations on the scale of hours at ecliptic latitudes 
$\leq 20$\degr. More than 95\% of all known main belt asteroids 
are found at 
inclinations $\leq 20$\degr, and studies have shown that the number 
counts of asteroids drop off by a factor of two from ecliptic latitudes 
of 0\degr{} to latitudes of 5\degr{} to 10\degr{} in the 
IR \citep{tedesco2005,ryan2009}. In order to detect the smallest, and 
thus faintest, asteroids in the \textit{Spitzer} data, multiple epochs 
were required such that images from two epochs could be subtracted to 
remove fixed objects and allow for multiple detections of a single 
asteroid in an image pair. To properly derive diameters and albedos from 
thermal data, 24~\micron{} fluxes are required as these 
fluxes are neither contaminated by reflected solar flux 
\citep[as is the case for 
wavelengths shorter than $\sim 5$~\micron, 
e.g.,][]{ryan2009,mueller2007}, nor on the Wien side of the 
asteroid spectral energy distribution (SED) where thermal fitting 
errors are highest \citep{ryan2010,harris2011}. Two \textit{Spitzer} 
surveys fulfilled these requirements -- the MIPSGAL and the 
Taurus surveys.

The MIPSGAL survey \citep[][\textit{Spitzer} Program ID 20597]{carey2009} 
was designed to survey 72 square 
degrees of the inner galactic plane at 24 and 70~\micron{} with the 
Multiband Imaging Photometer for Spitzer \citep[MIPS;][]{rieke04}. At 
low ecliptic latitudes (ecliptic latitudes from -1\degr{} to +14.2\degr), 
two epochs of MIPS Scan observations were taken with 
separations of 3 to 7 hours to allow for asteroid rejection from the 
final image stacks over a total ecliptic survey area of 29.4 square 
degrees. The MIPSGAL data were obtained in Cycle-2 of the 
\textit{Spitzer} cryogenic mission during the period 2005 September 
27--29~UT. 

The Taurus survey 
\citep[][\textit{Spitzer} Program ID Numbers 3584, 30816]{taurus_seminal} 
was designed to survey approximately 30 square degrees in the Taurus 
Molecular Cloud at 24 and 70~\micron{} with MIPS Scan observations. The 
Taurus Molecular Cloud is centered at $\sim$ 3\degr{} ecliptic latitude, 
and all data of this region were taken at separate epochs at 5 to 12 hour 
intervals to allow for asteroid rejection from the final image stacks. This 
region was observed twice in two different years to obtain the required 
stacked survey depth; the total asteroid survey area is equal to 
53.12 square degrees. The Taurus data were obtained in Cycle-1 and 
Cycle-3 of the \textit{Spitzer} cryogenic mission during the periods 
2005 February 27--March 2~UT and 2007 February 23--28 UT. 

\begin{figure}
\epsscale{1.}
\plotone{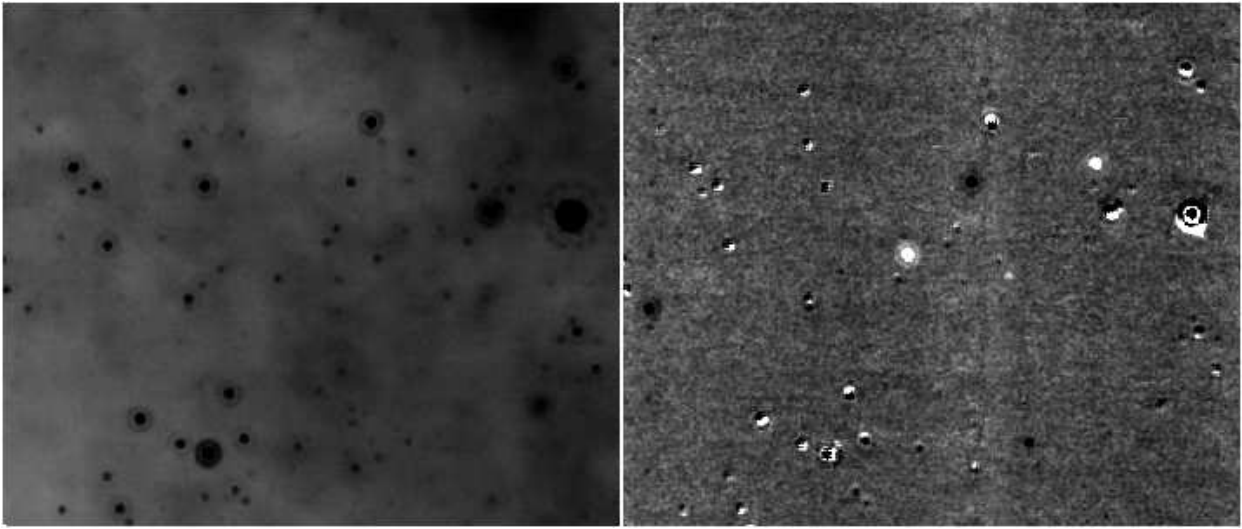}
\caption{Epoch subtraction example from the MIPSGAL survey. Each image has a field of view of 8.75 $\times$ 6.96 arcminutes. On the left is a single AOR (Request Key 15619072) and on the right is the image that results from subtracting this image from its image pair, Request Key 15644416. The three asteroids seen in this frame are San Juan, Fienga and 1321 T-2.}
\label{fig:subtraction_example}
\end{figure}

The MIPS 24~\micron{} band imager is a $128 \times 128$ pixel Si:As 
impurity band conduction detector with an effective wavelength of 
23.68~\micron{} with a native pixel scale of $2\farcs49 \times
2\farcs.60$. All 24~\micron{} data are diffraction limited. All data 
obtained in the MIPSGAL and Taurus programs utilized the MIPS 
Scan Astronomical Observing Template with a Fast Scan Rate resulting 
in a total integration time per pixel of 15.7~secs in each AOR mosaic. The image data files 
selected for our analysis from MIPSGAL consists of 42 Astronomical 
Observing Requests (AORs) (21 pairs) reprocessed with the MIPSGAL 
processing pipeline of \citet{Mizuno08}, except that 
asteroids were not masked out of the AOR mosaics. The image data 
files selected for our analysis from the Taurus surveys consists of 
30 AORs (15 pairs) processed by the data processing pipeline at 
the \textit{Spitzer Science Center}. The pairs of images which shared a 
common image center were registered utilizing the world 
coordinate system (WCS)
and differenced as illustrated 
in Figure~\ref{fig:subtraction_example}. Image subtraction allows for 
removal of fixed point sources and galactic background structure. 
MOPEX \citep{mm05} was employed in conjunction with 
single epoch uncertainty maps to produce 
point-spread function (PSF) fitting photometry of the positive and 
negative sources in each difference image 
consisting of object positions, fluxes and 1-sigma uncertainties 
in the point source fitted fluxes.Positive and negative source 
catalogs were constrained to only report objects detected with 
PSF chi-squared normalized by the degrees of freedom in the 
PSF fit greater than one, which results in only returning objects detected at a signal-to-noise ratio of five (5) or greater.
 
Due to small world coordinate system offsets between epochs, some 
fixed sources are detected in both the positive and negative source 
catalogs. To fully remove these sources from the asteroid candidate 
catalogs, the positive and negative catalogs are cross matched. Any 
object with a partner in the opposite catalog with a position within a 
radius of 1.5 pixels is rejected from the asteroids candidate catalogs. Each 
candidate catalog is also searched for false sources present in the 
data due to increased sensitivity in small regions with increased 
areal coverage in an AOR which appear with predictable offsets given 
the scan rate. These latent sources were also removed from the final 
candidate catalogs.   
 
Initial asteroid identification was performed utilizing known 
asteroids in the field. The JPL Horizons 
ISPY\footnote{http://ssc.spitzer.caltech.edu/warmmission/propkit/sso/horizons.pdf - Appendix 3} 
tool was queried on 22 January 2011 to produce lists of all 
known asteroids present in the MIPSGAL and Taurus images and that time, the 
Horizons database contained the orbital elements for 543,357 known 
asteroids. The ISPY tool requires input of observation time and image 
corners and produces a list of known asteroids which would be present in 
the field, along with predicted positions, the predicted apparent 
magnitude, and the instantaneous rates of change in Right Ascension (RA.) 
and Declination (Decl.) at the time of observation in arcseconds per hour. 
The observation time given for all ISPY queries was the observation time 
of the first BCD image in the AOR mosaic. ISPY queries were executed on an 
AOR basis, therefore for each subtracted image; two (2) ISPY queries were 
executed to predict the positions of the asteroids in each epoch. The 
predicted position at the start of an AOR, the AOR duration and the 
orbital rates are then convolved to define a 
search box for known asteroid candidates in each epoch. In 90\% of 
cases, only one object from the candidate asteroid catalog is present 
in the search box. We interpret this coincidence as a direct object 
match. In the cases where multiple candidates are detected within a 
search box, the predicted position of the known object and the matched 
candidates are output to a file for visual inspection and recovery. A 
list of non-detected asteroids in each field is also produced to 
estimate the completeness of the 24~\micron{} MIPS dataset. 

All matched known asteroids and their corresponding predicted and 
detected positions, fluxes and orbital parameters are reported in 
Table~\ref{table:flux_table}. Columns in the flux table 
include asteroid name, Request Key of associated observation, date and 
time of observation, predicted RA. and Decl., detected RA. and Decl., a 
flux data flag, 24~\micron{} flux and associated uncertainty, heliocentric 
distance ($r_{h}$) and \textit{Spitzer}-to-asteroid distance, phase 
angle ($\alpha$), and optical absolute magnitude ($H$).
The flux flag has a value of 1 for all objects except 
in cases where asteroid flux varies by $\gtsimeq 30$\% between two epochs, 
which is denoted with a flag of 2, or if an additional source such as 
a star is within 3\farcs75, which is denoted with a flag of 3. Five 
hundred eighty-eight (588) known asteroids were detected only once in 
the \textit{Spitzer} data, 1035 known asteroids were detected twice 
and 208 known asteroids were detected 3 or more times.

Eight (8) bright blended asteroid sources are present in the MIPSGAL 
and Taurus datasets. These asteroids are 103 Hera, 206 Hersilla, 2
33 Asterope, 318 Magdalena, 106 Dione, 1122 Neith, 283 Emma and 
2007 McCuskey. Due to the extreme brightness of these sources 
these sources are ``soft-saturated,'' and a single point 
source fitting result does not accurately measure the total flux from 
these objects. Fluxes for these objects were recovered 
determining the relative flux ratio between the model 24~\micron{} 
PSF central source and the first Airy ring and multiplying this 
ratio times asteroid fluxes in the first Airy ring. Positions 
in Table~\ref{table:flux_table} are the nominal positions of the 
saturated PSF center and the uncertainty in the reported fluxes is 
assumed to be 15\%. A flux flag 
value of 4 in Table~\ref{table:flux_table} is used to denote the 
instances where the reported fluxes for these objects are reported 
from a saturated source.

Asteroid candidates with no association with known asteroids 
were not used for further diameter and albedo analysis. Estimating 
diameters and albedos for these objects is highly uncertain without 
the availability of optical absolute magnitudes, and derivation of 
orbits for these objects is problematic with only two epochs of data as 
acceleration vectors cannot be derived from the positional data.

\section{ALBEDO DETERMINATION}\label{sec:alb_det}

We used the Near Earth Asteroid Thermal Model 
\citep[NEATM;][]{har98} to determine the rotationally averaged 
diameters and albedos of 
known asteroids in our MIPSGAL and Taurus samples. The NEATM 
relies upon a 
basic radiometric method to determine both the diameter and albedo of an 
asteroid \citep[for a complete discussion, see][]{ryan2010}. NEATM 
assumes balance between incident radiation and emitted radiation, 
where the emitted radiation has two components; a reflected and a 
thermal component. The reflected component has approximately same 
spectral energy distribution (SED) as the incident radiation; i.e., 
the reflected component is dominant in the optical and peaks in V band 
commensurate with the spectral region in which the sun emits the 
greatest flux. The reflected asteroid flux is proportional to the 
diameter of the body, $D (km)$ and the geometric albedo, $p_{V}$. To 
maintain energy balance the thermal flux is proportional to the amount 
of incident flux which is not reflected.


However, asteroids do not maintain one single body temperature, T(K), 
rather there is a temperature distribution across the surface which is 
then observed in the mid-IR. The NEATM utilizes an assumed temperature 
distribution to model the total IR flux, which is related to $p_{V}$. 
The temperature distribution utilized by NEATM is: 

\begin{equation}
T_{NEATM}(\phi, \theta)= \left [ \frac{(1-A)S_\odot}{\eta r_{h}^{2} \epsilon 
\sigma} \right]^{\frac{1}{4}}  (cos \phi)^{\frac{1}{4}} (cos \theta)^{\frac{1}{4}} 
\label{eqn:neatm_e}
\end{equation} 

\noindent where the temperature, $T$ is in Kelvin, $A$ is the geometric 
Bond albedo, S$_{\odot}$ is the solar constant (W~m$^{-2}$), r$_{h}$ 
is the heliocentric distance (AU), $\epsilon$ is the emissivity of the 
object (assumed to be 0.9), $\sigma$ is the Stefan-Boltzmann constant, 
$\eta$ is the beaming parameter, $\phi$ is the latitude, and 
$\theta$ is longitude of the coordinate grid superposed on the 
asteroid. The derived temperatures are not phase angle dependent in the 
NEATM approach \citep{har98}.

In the NEATM temperature distribution, $\eta$, the beaming parameter 
is utilized as a variable to characterize both shape and 
thermal inertia. In an ideal case where an asteroid is a perfect 
sphere with zero thermal inertia, $\eta$ equals unity. Only one thermal 
photometric measurement is available from the 24~\micron{} measurements; 
therefore, NEATM was run with a fixed beaming parameter of $\eta = 
1.07$. This value of $\eta$ was selected by averaging the value of 
$\eta$ for 1584 main belt, Hilda and Trojan asteroids observed by IRAS 
and/or MSX \citep[i.e.,][]{ryan2010}. In addition we adopt a 
value for the emissivity ($\epsilon$) of 0.9, a value appropriate 
for rock \citep{morrison1973}, and a phase slope 
parameter ($G$) of 0.15 when computing the asteroid diameter and 
albedo. To compute 
the geometric albedo and thus the temperature distribution on the 
illuminated face of the asteroid, one must anchor solutions to an 
optical data point. We utilized optical absolute magnitudes ($H$) from the 
Minor Planet Center \footnote{www.cfa.harvard.edu/iau/mpc.html}(MPC) 
for the purposes of our solutions. 

The validity of our thermal models to compute rotationally 
averaged parameters of asteroids is robust and as yielded model 
albedo and diameters that are consistent with radar and 
occultation measurements of many tens of 
asteroids \cite[e.g.,][]{ryan2010}.

All mean albedo and mean diameter solutions reported in 
Table~\ref{table:sighting_solns} are derived from Monte Carlo modeling 
for each asteroid per sighting. A 500 data point distribution was 
created for each object 
observation such that the mean flux was equal to the flux measured by 
MOPEX and the standard deviation of the distribution was equivalent to 
the uncertainties in the flux measurement. These flux points were then 
used in conjunction the known orbital parameters and the $H$ 
magnitude to produce albedo and diameter fitting results. 
In this Monte Carlo modelling, the optical absolute magnitude ($H$)
was also varied by up to 0.2 magnitudes; equal to the mean offset in 
asteroid absolute magnitudes as derived from the MPC and the Asteroid 
Orbital Elements Database 
\citep[ASTORB\footnote{ftp://ftp.lowell.edu/pub/elgb/astorb.html};][]{bowell1989}.
Due to the wide width ($\Delta\lambda = 4.7$~\micron) of the MIPS 24~\micron{} channel, a color correction is 
also required to accurately fit the albedo and diameter.  Our 
implementation of NEATM applies color corrections iteratively, such 
that a color correction is applied to the model asteroid flux with each refinement of the albedo 
\citep{ryan2010}. Instead of using the subsolar temperature for the 
color correction, we calculate the mean of the temperature 
distribution for the application of the color correction, as described 
in \citet{ryan2010}. The standard deviation of the albedos and 
diameters listed in Table~\ref{table:sighting_solns} are the 
1-sigma statistical uncertainty ($\pm$) in the results added in 
quadrature with a $\pm 2\%$ error in 
the absolute calibration of MIPS 24~\micron{} data \citep{Gordon05}.  
Results reported in Table~\ref{table:sighting_solns} are sorted 
by AOR Request Key and asteroid name/provisional designation allowing 
for direct matching of results with input data by line in 
Table~\ref{table:flux_table}. Table~\ref{table:mean_solns} is sorted in 
alphanumeric order and reports the mean albedo and diameter and 
associated 1-sigma uncertainties for 1831 asteroids, as well as the 
number of sightings used to arrive at these solutions.

\section{THERMAL MODELLING RESULTS}\label{sec:modeling_results}

\subsection{Albedo and Diameter Properties/Validity}

Prior observations and thermal model fits of the nine brightest sources, 
103 Hera, 206 Hersilla, 233 Asterope, 318 Madgalena, 106 Dione, 
1122 Neith, 283 Emma, 2007 McCuskey and 106 Aethusa derived 
using the 
NEATM and \textit{IRAS} or \textit{MSX} photometry were compared to 
those obtained from the MIPS photometry. Table~\ref{table:big_asteroids} 
summarizes the albedos and diameters computed from the MIPS 24~\micron{} 
data, and their \textit{IRAS} or \textit{MSX} derived NEATM albedo, 
diameter and beaming parameter \citep{ryan2010} and an occultation 
diameter if available from \citet{Dunham09}. The diameter estimates from 
MIPSGAL photometry are within the uncertainties of those derived from 
fitting the SEDs produced by minimum of three wavelength 
specific fluxes from the \textit{IRAS} and \textit{MSX} surveys for 
all the objects. The thermal model solutions for the three asteroids with occultation 
derived diameters also match within 1$\%$. This overlap in the diameter 
estimates suggests that the MIPSGAL and Taurus 24~\micron{} solutions 
are robust, which is not surprising as we are observing thermal 
emission from asteroids at the peak of, or
on the Rayleigh-Jeans tail of the SED. However, a slight variation in 
the mean diameters is present. This small spread in the distribution can be 
attributed to the use of a single mean beaming parameter of $\eta = 1.07$ for 
all asteroids in the MIPS data. This value of $\eta$ must be 
used as there is insufficient photometry 
to allow for independent fits of albedo, diameter, and beaming 
parameter simultaneously. \citet{walkercohen2002} in their analysis
of \textit{IRAS} Low Resolution Spectrometer (LRS) SEDs derive a mean
$\eta = 0.98$ with a one sigma uncertainty of 0.08, commensurate
with the value we adopt in our work.  The 
agreement between \textit{IRAS} and \textit{Spitzer} results for these 
diameters is evidence that utilization of a single beaming parameter is 
appropriate for a bulk treatment of main belt asteroids. 

\textit{Spitzer} 70~\micron{} data likely would provide an additional 
constraint on the derived characteristics of the sample asteroids. 
Unfortunately half of 
the 70~\micron{} array malfunctioned, and the default MIPS scan AOT 
used for these observations leaves large gaps in the mosaics, resulting 
in a striped 70~\micron{} mosaic wherein useful data only exists for 
half of the areal coverage of a 24~\micron{} mosaic. This poor 
coverage coupled with the low sensitivity of 70~\micron{} fast scan 
maps, which is insufficient to recover 90 km asteroids, led us to discard these data from our analysis. 

Use of a single beaming parameter based on \textit{IRAS} 
and \textit{MSX} results assumes that small and large asteroid bodies 
have similar surface roughnesses and thermal inertia. For main belt asteroids 
with diameters $> 20$~km, thermal inertia and diameter are inversely 
proportional \citep{delbo2009} and this net effect would drive the 
beaming parameter to larger values. A similar effect of increasing thermal inertia with decreased size is noted within the Near Earth Asteroid population as well \citep{delbo2007}. If a beaming parameter of 1.31 (equal 
to the mean from \textit{IRAS} and \textit{MSX} plus a 1-sigma standard 
deviation) is assumed for the 24~\micron{} thermal modeling, the albedos 
differ from those reported here by $\leq 10$\% and the albedos differ 
by $\sim 5$\%. If a beaming parameter of 0.77 (equal to the mean from 
\textit{IRAS} and \textit{MSX} minus a 1-sigma standard deviation) is 
used for the 24~\micron{} thermal modelling, the albedos differ from 
those reported here by $\sim30$\% and the offsets are not systematic.

An additional source of albedo and diameter uncertainty is 
related to the reliability of the optical absolute magnitudes provided by 
the MPC. A systematic color dependent offset was found between apparent 
V~band magnitudes calculated using ASTORB \citep{bowell1989} orbital 
elements and absolute magnitudes and the synthetic V~band photometry 
derived from the Sloan Digital Sky Survey \citep[SDSS;][]{juric02} of 
order 0.34 and 0.44 magnitude respectively for the blue and red populations 
of asteroids. This magnitude discrepancy is lessened to a 0.2 magnitude 
offset when MPC absolute magnitudes are used to derive a projected 
apparent magnitude. While a systematic offset in projected apparent 
magnitude could be the result of the SDSS using only 
two-body computations 
to calculate $r_{h}$, geocentric distance ($\Delta$) and phase 
angle, the relative offset of 0.1 magnitudes between red and blue 
objects is accounted for in our Monte Carlo modelling allowance of $\pm$~0.2 magnitude variation 
in the optical absolute magnitude $H$.

\subsection{Completeness}

To place the albedos and diameters in the MIPSGAL and Taurus catalogs 
in context, the effects of optical and IR completeness must be 
considered. To assess the completeness of optical asteroid surveys, we 
assume that they are complete to a V magnitude of 21.5, 
commensurate with the 95\% completeness limit from 
the SDSS \citep{juric02} and other surveys such as the  
Sub-Kilometer Asteroid Diameter Survey \citep{gladman09} and Spacewatch 
\citep{larsen07}. Assuming that an asteroid will be detected at opposition 
by one of a number of surveys, we utilize the relation 
$m_{V}=H + 5 log [r_{h}(r_{h}-1)],$ and calculate the completeness limits 
in terms of $H$ in each of the four main belt asteroids zones as 
defined in \citet{zellner1975}, adopting opposition and a heliocentric 
distance which corresponds to the outer semimajor axis range of each 
respective zone. These values range from absolute magnitudes of 18.6 in the inner main belt to 16.57 in the outer main belt.
Unfortunately, many asteroid surveys are pencil beam surveys which do not cover the full sky, therefore we estimate the full sky completeness of asteroid surveys utilizing the H magnitude distributions from the Minor Planet Center. We assume that all optical surveys are complete in each main belt region to the magnitude bin which contains the highest number of asteroid sources, and report those H magnitudes in Table~\ref{table:completeness_table}.
Values in Table~\ref{table:completeness_table} reflect 
both the completeness in terms of $H$  and 
diameters assuming a mean asteroid geometric albedo $p_{V}= 0.02$ commensurate with the darkest observed asteroid albedos from any survey.

Queries of the Horizons database via the ISPY tool predict a total 
number of 7598 asteroid appearances for 3429 individual 
asteroids in the MIPSGAL and Taurus surveys. The catalog produced in 
this work contains 3486 sightings of 1831 individual asteroids, resulting 
in an overall object detection rate of $\sim 53$\%. There are three 
possible causes for this low recovery rate: rates of asteroid motion too 
low for the detection of movement between epochs; high rates of asteroid 
motion during a single AOR; and the mid-IR sensitivity 
completeness cut off. 
Those asteroids whose rates of motion would make them appear as fixed 
targets in the two epoch MIPSGAL data are Centaurs or Kuiper Belt 
objects. From the instantaneous rates of change in RA. and Decl. 
provided via the ISPY query, 32 objects are found to have rates of 
motion that would be insufficient for two epoch detection via the 
subtraction method for the shortest epoch separation of 3 
hours. Near Earth 
asteroids (NEAs) are objects which move at such high rates that 
they may not be matched in an AOR due to smearing of the flux along the 
direction of motion. The rates of motion required for an asteroid 
source to move 1.2\arcsec{} (half of a native MIPS pixels) in an individual 
5~sec BCD and a 15~sec stacked mosaic 
are $\sim 1464$ arcseconds per hour and 
$\sim 293$ arcseconds per hour, respectively. The asteroid 2002 AL14 has 
the greatest instantaneous predicted rate of motion of 186 arcseconds per 
hour in this survey data and was recovered in all three epochs where 
sightings were predicted. Therefore, the expected losses due to asteroid 
motion are biased towards non-detection of slowly moving objects in both 
the MIPSGAL and Taurus datasets and the losses are less than 1\% in total.

To assess the completeness of the 24~\micron{} data, synthetic sources 
were added to single epoch MIPSGAL and Taurus AORs which were subsequently 
subtracted following the data analysis techniques described in 
Section~\ref{sec:observ_mips}. The 90\% completeness limit was found to 
be 2~mJy for the MIPSGAL survey. We adopt this as the 24~\micron{} 
completeness limit of both the MIPSGAL and Taurus surveys, although the 
Taurus survey is complete to 1.5~mJy due to a lack of extended background 
emission when compared to the MIPSGAL regions. At 24~\micron, the fluxes 
of asteroids are most highly dependent upon diameter, not albedo. This 
enables completeness estimates as a function of diameter to be derived
within the same zones utilized to analyze the optical completeness as 
reported in Table~\ref{table:completeness_table}. These diameters were 
also calculated via the NEATM flux distribution with $\eta= 1.07$, assuming 
an object was observed at opposition (thus at a phase angle of zero 
degrees), and that the \textit{Spitzer}-to-asteroid distance was 1~AU 
less than the heliocentric distance.

The asteroid completeness limits derived from the optical and the 
MIPSGAL and Taurus surveys are compatible; however, for all subsequent 
analysis, objects were removed which had fluxes less than the 
24~\micron{} completeness limit of 2 mJy or if they had an $H$ 
magnitude greater than the optical completeness limit in their region. 
This constraint de-biases the \textit{Spitzer} sample and  resulted 
in the removal of 289 objects from the combined MIPSGAL and Taurus 
catalogs in the subsequent analysis; 280 for having $H$ values 
less than the optical completeness, 4 for having fluxes less than the 24~\micron{} completeness limit, and 5 for having both $H$ values less than the optical completeness \textit{and} fluxes less than
the 24~\micron{} completeness limit.

\begin{figure}
\epsscale{1.15}
\plottwo{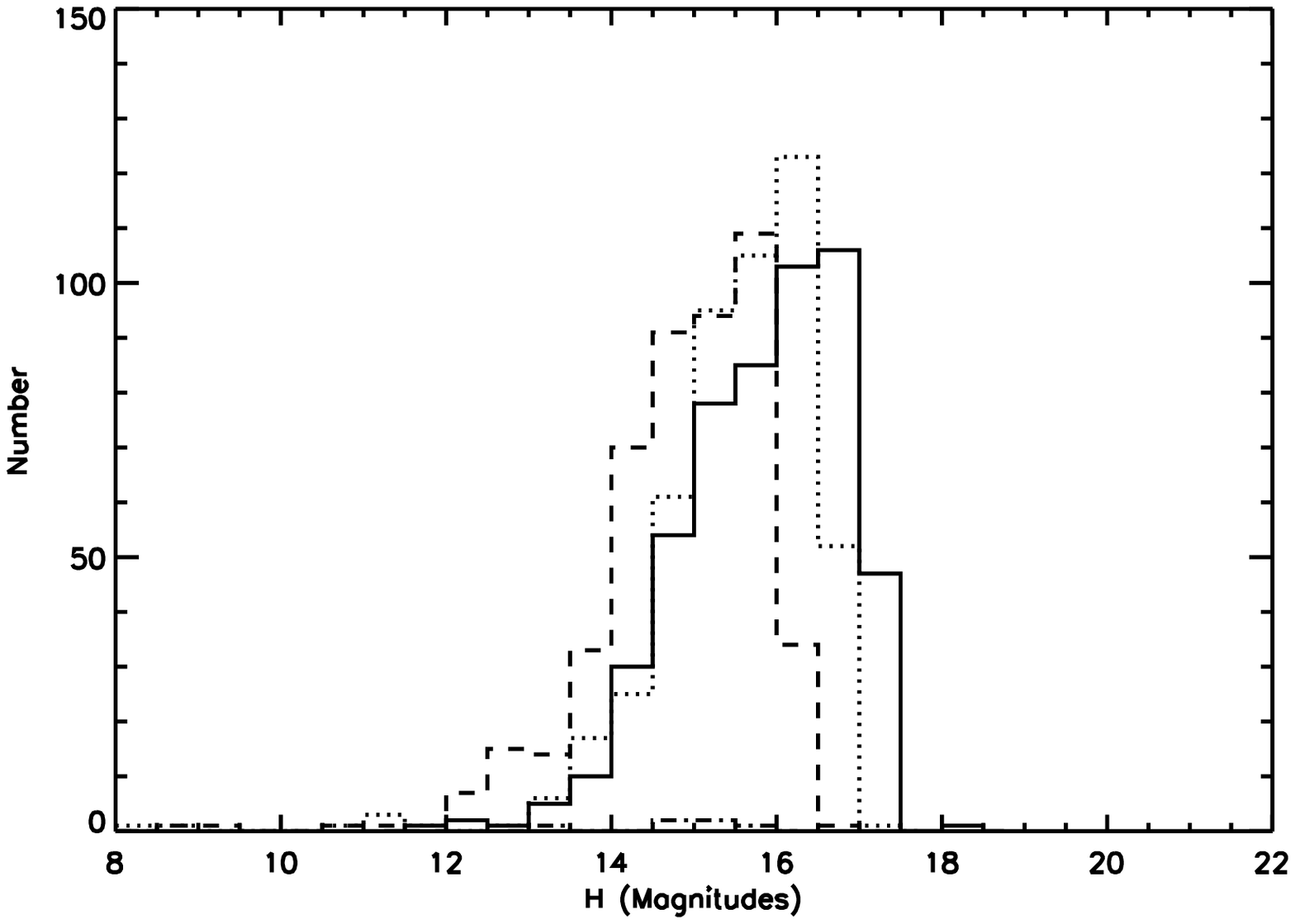}{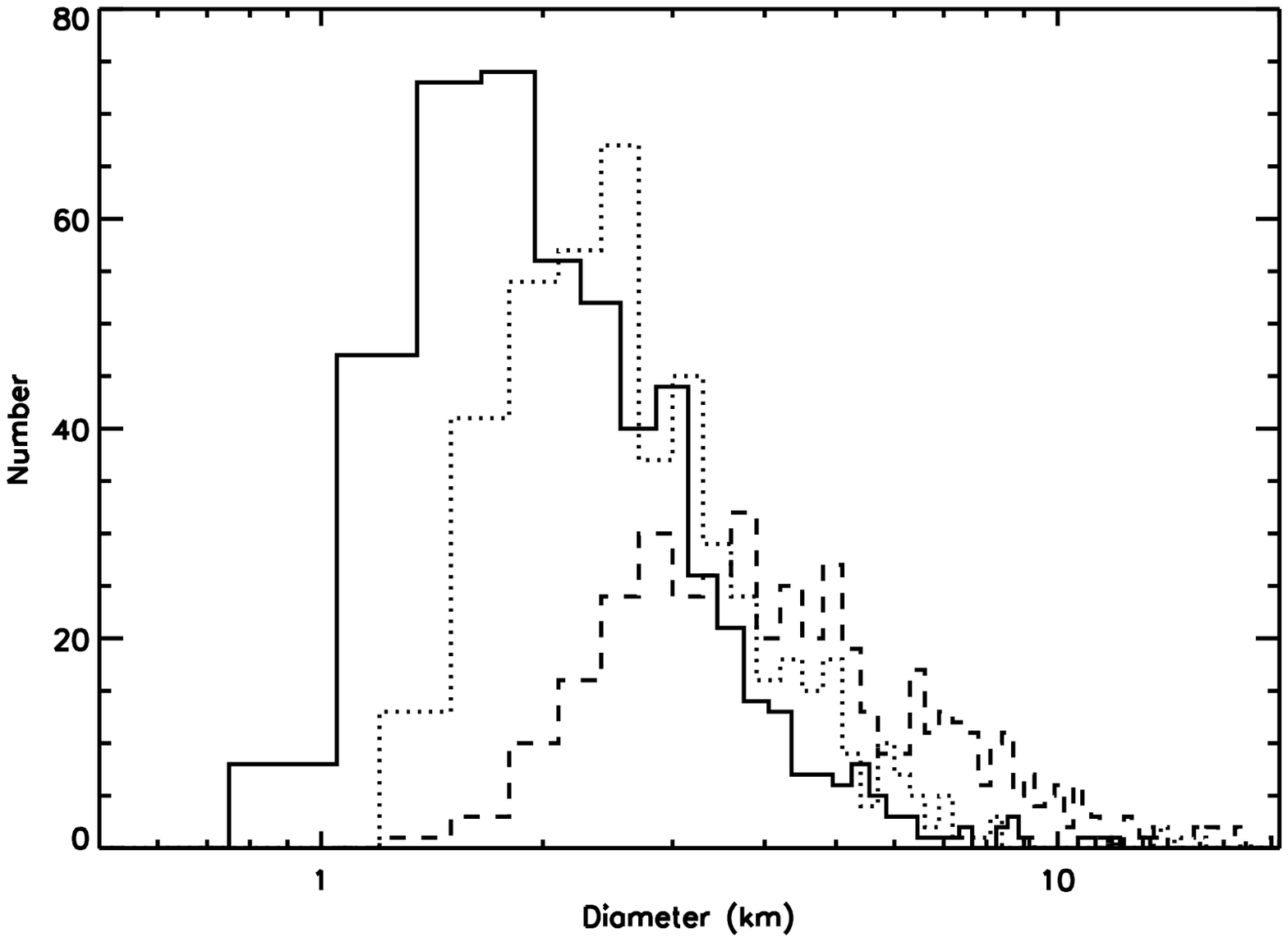}
\caption{\textit{Left:}$H$ magnitude distributions of \textit{Spitzer} detected asteroids: the solid line corresponds to the \citet{zellner1975} Main Belt Asteroid (MBA) I region (2.06 $<$ a $\le$ 2.5), the dotted line corresponds to MBAII (2.5 $<$ a $\le$2.82), the dashed line corresponds to MBAIII (2.82 $<$ a $\le$3.27). \textit{Right:} Diameter distributions of the \textit{Spitzer} detected asteroids in the same regions as the H magnitude distributions. }
\label{fig:hmag_dists}
\end{figure}

It is useful to compare the relative completeness of the 
MIPSGAL and Taurus surveys to the NEOWISE survey \citep{wise2011}. 
With a mid-IR completeness limit of the NEOWISE survey currently 
unavailable \citep[cf.,][]{masiero2011}, we utilize the $H$ 
magnitude distributions as a function of semi-major axis 
to compare relative completeness between the NEOWISE, MIPSGAL, 
and Taurus surveys. In the inner main belt, the $H$ magnitudes of 
the NEOWISE 
survey peak at $H \simeq 15.5$ \citep[Fig.~8 of][]{wise2011}, whereas 
the mean $H$ magnitude of the MIPSGAL and Taurus catalogs in the inner 
main belt is 15.8~mag as illustrated in Figure~\ref{fig:hmag_dists} once sources with H magnitudes greater than the optical completeness 
limit are removed. This 
offset of $\sim 0.3$ magnitudes between the NEOWISE and MIPSGAL and Taurus 
surveys is consistent in all semimajor axis zones. This offset translates 
roughly into a diameter difference of 0.5 ~km at any given albedo indicating 
that the \textit{Spitzer} data is detecting asteroids at least 
0.5~km \textit{smaller} than NEOWISE survey at any given region of the 
asteroid belt. 

This offset is confirmed also by a simple estimate of the 
asteroid diameters which can be detected by the WISE mission 
\citep{wright2010}. Assuming a 5-$\sigma$ limiting flux of 10~mJy at 
22~\micron, a beaming parameter, $\eta = 1.07$, geometric albedo 
$p_{V} = 0.14$, and an asteroid observed at opposition (phase 
angle $\alpha = 0$) at a heliocentric distance of 2.5~AU and a 
geocentric distance ($\Delta$) of 1.5~AU, WISE can only detect 
asteroids with diameters $\gtsimeq 1.65$~km, whereas under these s
ame orbital assumptions and a flux completeness limit of 2~mJy, 
our \textit{Spitzer} data is sensitive to asteroids with diameters 
$D \gtsimeq 0.79$~km.

\subsection{Albedo Catalog Comparision}\label{sec:albedo_comparison}

\begin{figure}
\epsscale{0.8}
\plotone{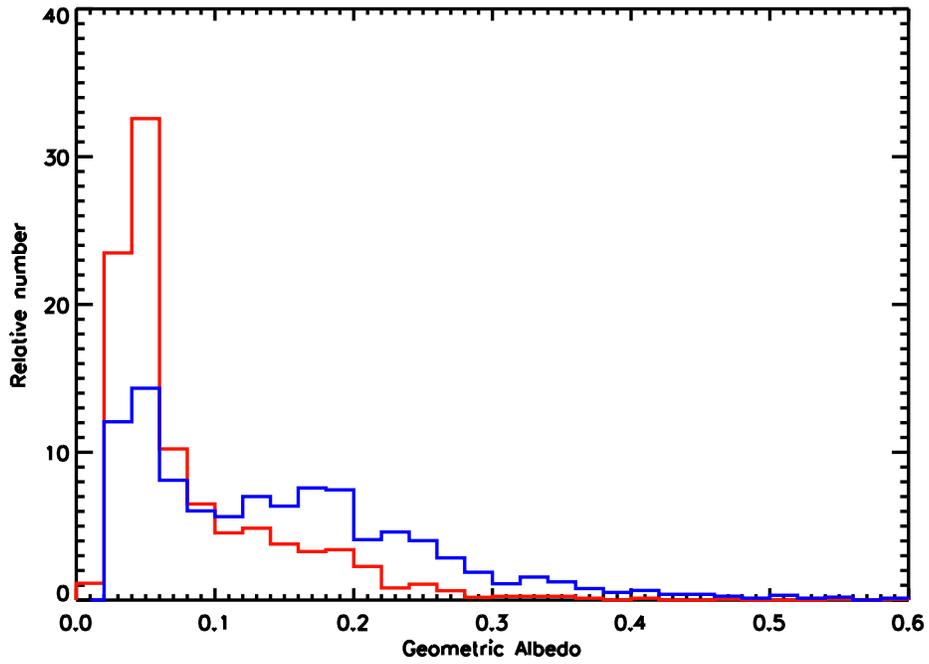}
\caption{Albedo distribution for asteroids in the \textit{IRAS} and \textit{MSX} catalogs of \citet{ryan2010} plotted in red and the albedo distribution for small Main Belt asteroids from the MIPSGAL and Taurus surveys shown in blue. The y-axis is percent of the total sample in each bin - addition of all y-values will equal 1 (100\%)}
\label{fig:IRASvMIPSGAL}
\end{figure}

The albedo distribution histogram from the MIPSGAL and Taurus surveys 
is presented in Figure~\ref{fig:IRASvMIPSGAL}.  The albedo distribution 
derived for small asteroids is more diverse than the albedo distribution 
for large asteroids derived from \textit{IRAS} and \textit{MSX} data. The 
mean albedo for the complete \textit{Spitzer} sample is 0.147
with a sample standard deviation of 0.104, whereas the mean albedo for 
the \textit{IRAS} and \textit{MSX} sample of 1584 objects in \citet{ryan2010} 
is $p_{V} = 0.081$ with a sample standard deviation of 0.064.
To test if these small and large asteroid albedo distributions were 
selected from the same parent distribution, we performed a 
Komolgorov-Smirnov (K-S) test which rejected this hypothesis at the 99.99 
percent level. We also performed a Wilcoxon-Mann-Whitney test to 
determine if these two albedo distributions were selected from a 
population with the same mean and that hypothesis was also rejected at 
the 99.99 percent level.

Studies of space weathering 
\citep[][and references therein]{Nesvorny2005,clark2002} indicate 
that young collisional fragments have different colors or higher albedos than old asteroids 
which have been subjected to solar wind exposure or micrometeorite 
impacts. Lunar space weathering causes microscopic melting 
of the surfaces and the formation of agglutinates. Impact melt causes
submicroscopic metallic iron on the surfaces of lunar 
regolith particles (which in general are highly
regular in shape, i.e., spherical) in cases where iron-bearing assemblages 
are extant on the surface \citep{keller1998, keller1999, pieters2000}. The 
net effect is a reddened slope and decreased albedo with increasing 
exposure time. This lunar-type space weathering is assumed to modify 
the surfaces of asteroids as well, as minimal lunar-type space weathering 
is needed to match ordinary chondrite spectra to the spectra of 
S-type asteroids \citep{hapke2000, hapke2001}. However, analysis 
of asteroid regolith from the Hayabusa 25143 Itokawa sample return 
mission \citep{noguchi2011,tsuchiyama2011} seems to suggest that 
asteroid the optical properties of asteroid surfaces may be altered by 
a combination of radiation-induced amorphization in addition to in situ 
reduction of regolith iron by solar wind irradiation. Whether or not the 
properties of the regolith dust from asteroid Itokawa are representative 
of all asteroids in general awaits further in situ sample return 
confirmation \citep{brucato2010, lauretta2010}.

Space weathering likely affects the albedo diversity we observe 
in our sample. Such diversity can be explained if lunar-type space 
weathering effects, comprising micrometeor bombardment and solar wind 
irradiation, dominate. The distribution peak for carbonaceous asteroids 
is at a similar albedo for both large and small asteroids as there is 
insufficient iron bearing minerals for this iron sputtering on regolith 
particles. The high albedo tail of the small asteroid population also 
exhibits a greater diversity than the large asteroid population. A 
complication to this interpretation is that lunar-type space weathering 
has only been well studied with asteroids and mineralogies characteristic 
of the S- taxonomic type in observations and lab studies; the effects of 
solar wind exposure on compositions similar to C-type asteroids have 
not been the subject of laboratory investigation. Whether or not space 
weathering effects are germane is discussed further in 
Section~\ref{sec:family_trends}, where characteristics of individual 
dynamical families within the main belt are compared.

\subsection{Albedo gradient across Main Belt}

\begin{figure}
\epsscale{1.0}
\plotone{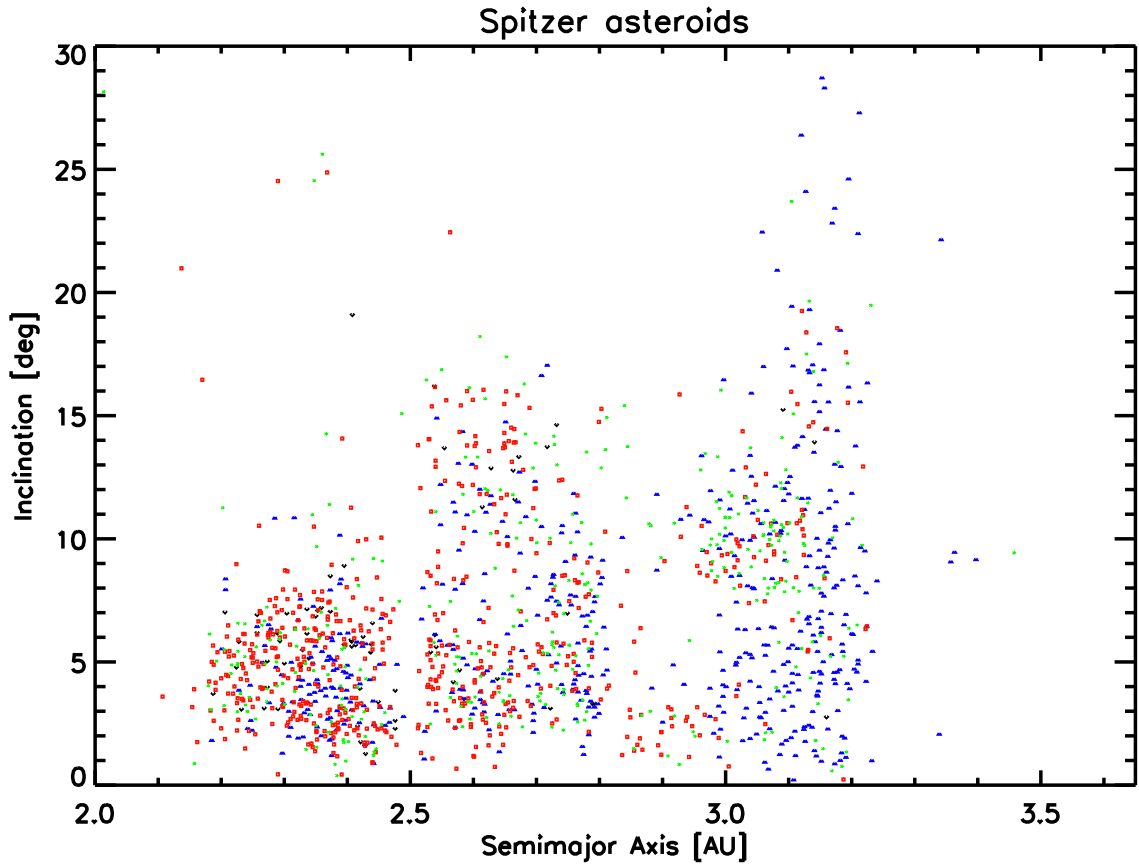}
\caption{Orbital Element distribution of asteroids from the \textit{Spitzer} MIPSGAL and Taurus surveys. Blue symbols correspond to asteroids with C-type albedos, green symbols correspond to asteroids with X-type albedos, and red symbols correspond to asteroids with S-type albedos.  The panel includes all asteroids in the \textit{Spitzer} catalogs.}
\label{fig:color_coded_albedos}
\end{figure}

A population of small thermally unaltered asteroids should exist in the 
inner main belt if $^{26}$Al melting models are 
correct \citet{McSween2002,grimmmcsween93}. To critically 
examine this hypothesis, we have analyzed the albedo-orbital distribution 
of asteroids in the MIPSGAL and Taurus surveys, 
Figure~\ref{fig:color_coded_albedos}, where the bulk albedo distribution 
of asteroids is color coded by albedo. The bulk albedo distribution, 
Figure~\ref{fig:color_coded_albedos} (\textit{left}) can be contaminated 
by dynamical family members; for example a single family of many small 
S-type fragments can make the outer main belt appear silicate rich. To 
determine the effects of dynamical families on the heliocentric distribution 
of albedo types, our MIPSGAL and Taurus catalogs were cross referenced 
with the Dynamical Family Catalog of \citet{nesvornycat} which utilized 
the proper elements for 293,368 asteroids to discriminate family memberships 
for 55 dynamical families. Of these 55 dynamical families, 47 are 
represented in our data and only eight families have more than 20 members 
in our combined MIPSGAL and Taurus albedo catalogs. 
Figure~\ref{fig:color_coded_albedos} (\textit{right}) shows the resultant 
bulk albedo-orbital distribution if these dynamical families are removed.

\begin{figure}
\epsscale{1.15}
\plottwo{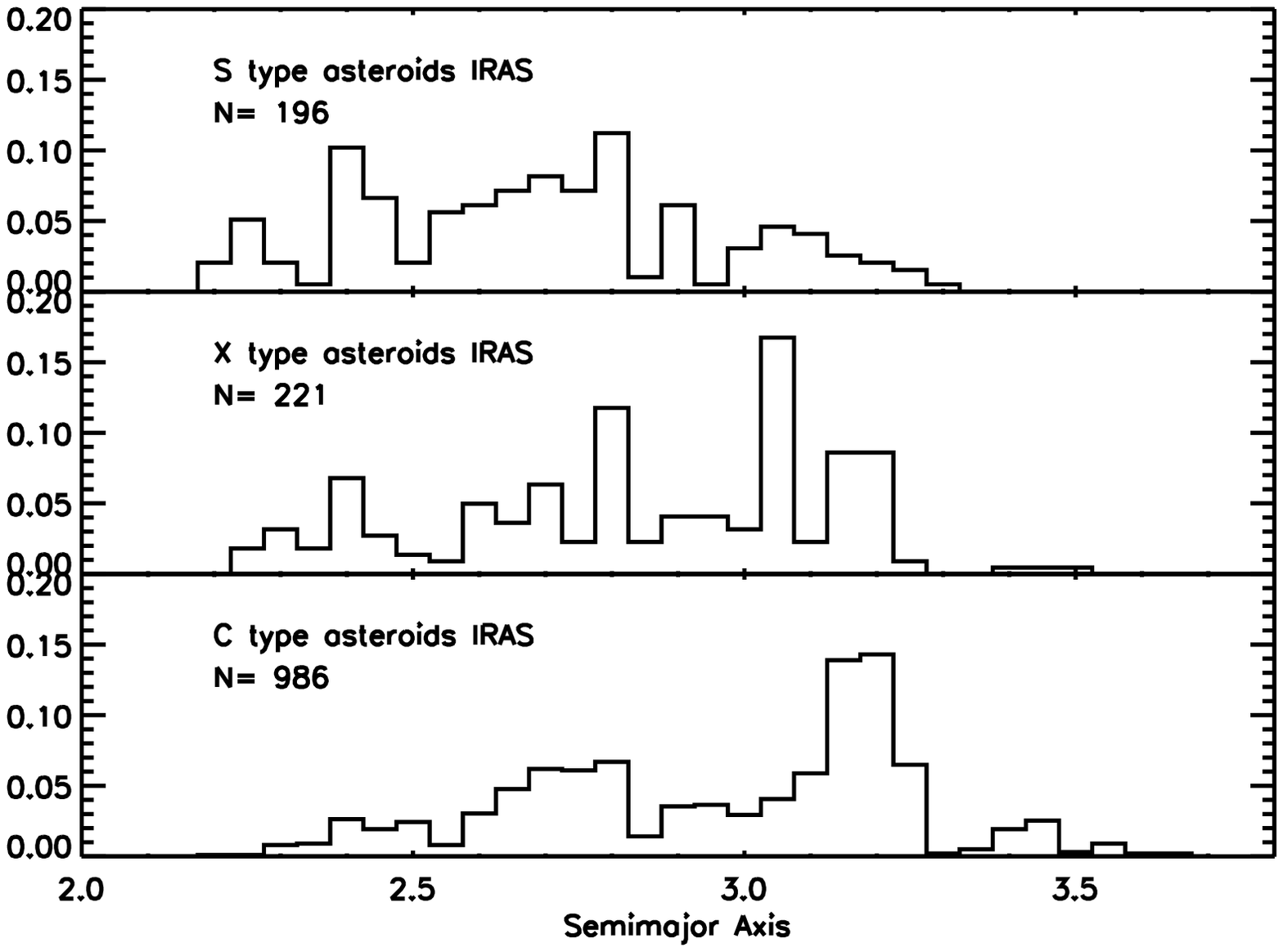}{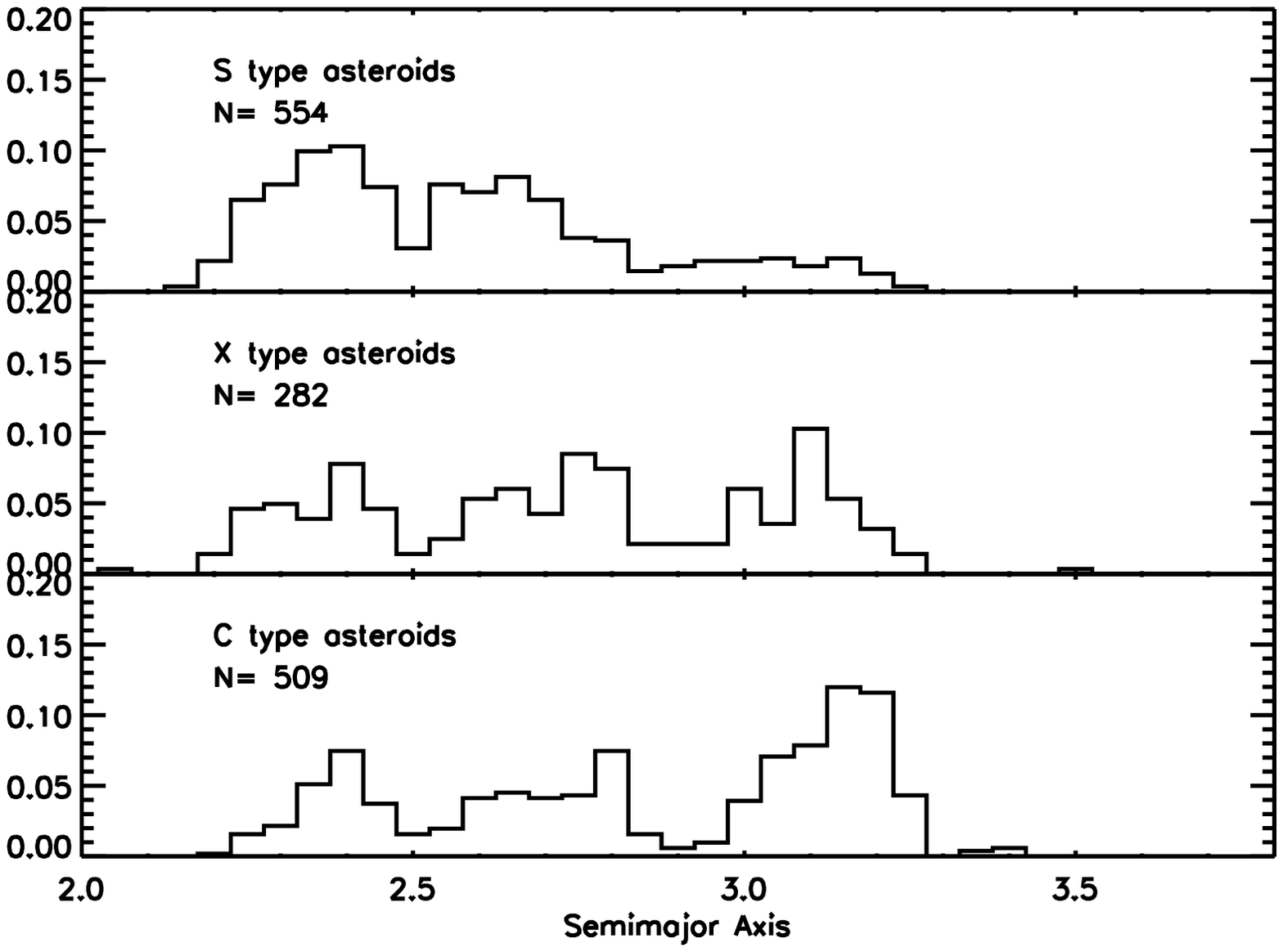}
\caption{Heliocentric distance distribution of C-, X-, and S-type asteroids from the \textit{IRAS} and \textit{MSX} surveys (\textit{left} panel) and this  \textit{Spitzer} work (\textit{right} panel). The y-axis is the total percent of the taxonomic type in each bin of width 0.05 AU - addition of all y-values will equal 1.0}
\label{fig:semimajor_axis_dist_all}
\end{figure}

\begin{figure}
\epsscale{1.15}
\plottwo{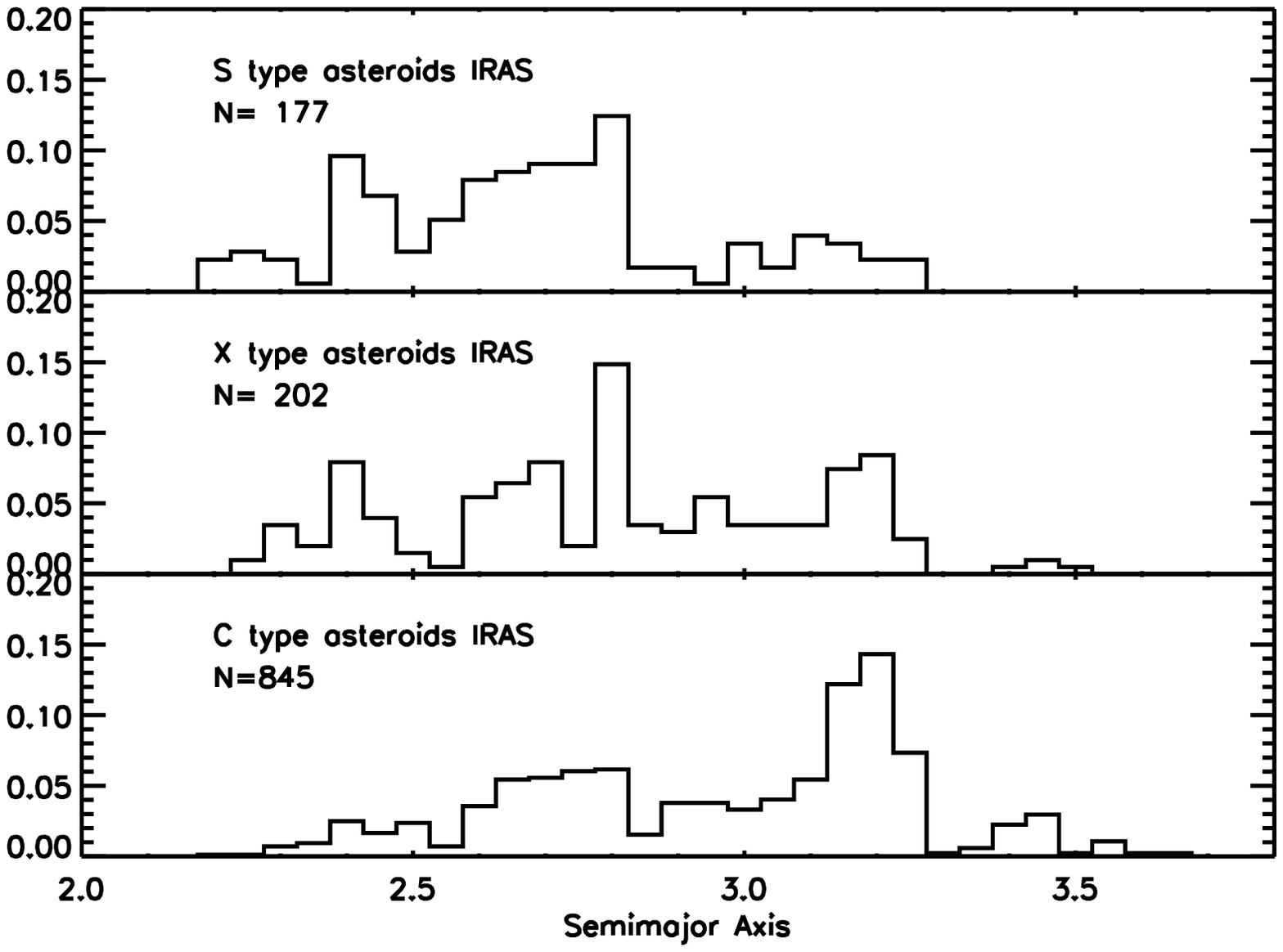}{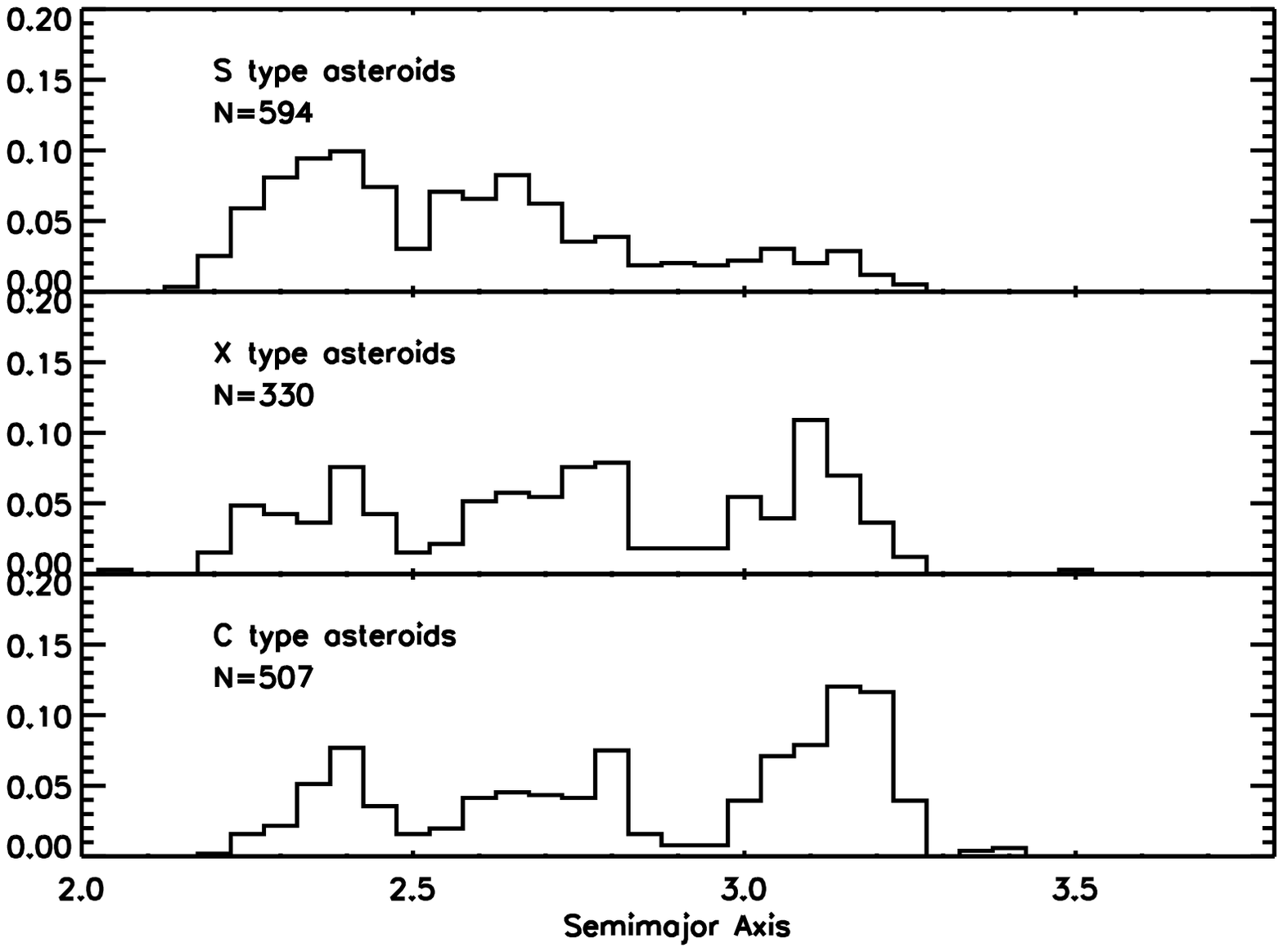}
\caption{Heliocentric distance distribution of C-, X-, and S-type asteroids with all family members removed from the \textit{IRAS} and \textit{MSX} surveys (\textit{left} panel, and  \textit{Spitzer} surveys on the \textit{right}. The y-axis is the total percent of the taxonomic type in each bin of width 0.05 AU - addition of all y-values will equal 1.0}
\label{fig:semimajor_axis_dist_fam_removed}
\end{figure}

To compare the albedo distribution of small main belt asteroids in this 
dataset to those large asteroids detected by \textit{IRAS} and 
\textit{MSX}, we utilize the albedo definitions of S, X and C complex 
asteroids from \citet{ryan2010} where C-types have $p_{V} \le 0.08$, 
X-types are described by a geometric albedo 0.08 $< p_{V} \le$ 0.15, 
and S-types span the range of geometric albedos 0.15 $< p_{V} \le$ 0.35. 
The semimajor axis distribution of each classification from 
\textit{IRAS} and MIPSGAL/Taurus is displayed in 
Figure~\ref{fig:semimajor_axis_dist_all}.  Though the semimajor 
axis distributions of S- and X-type asteroids appear similar between 
the \textit{IRAS} and \textit{Spitzer} surveys, the C-type distributions 
show a marked enhancement within the inner main belt. Removal of 
dynamical families, Figure~\ref{fig:semimajor_axis_dist_fam_removed}, 
does not markedly change in the overall semimajor axis distribution of 
taxonomic type. Twenty-two percent (22\%) of all small dark ($p_{V} \le 0.08$) , presumably 
carbonaceous, asteroids reside in the inner main belt 
(2 AU $\le$ a $\le$ 2.5 AU). An enhancement of small C-type asteroids 
in the inner main belt is commensurate with the study of 
\citet{carvano2010} who find that the distribution of small C- and 
X-type asteroids observed by the SDSS are fairly evenly distributed as 
a function of semimajor axis. 

\subsection{Dynamical family albedos\label{sec:family_trends}}

\begin{figure}
\epsscale{1.15}
\plottwo{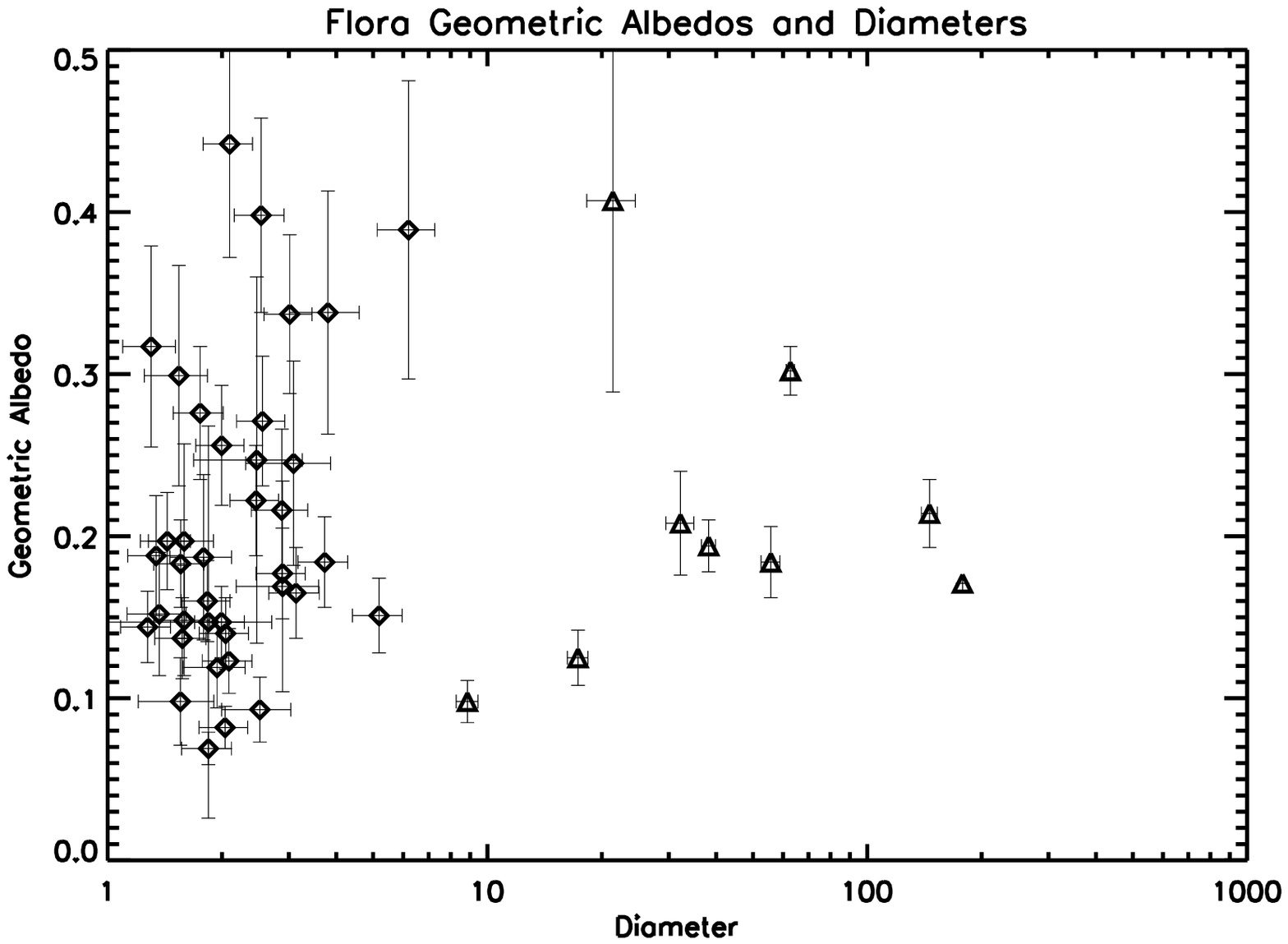}{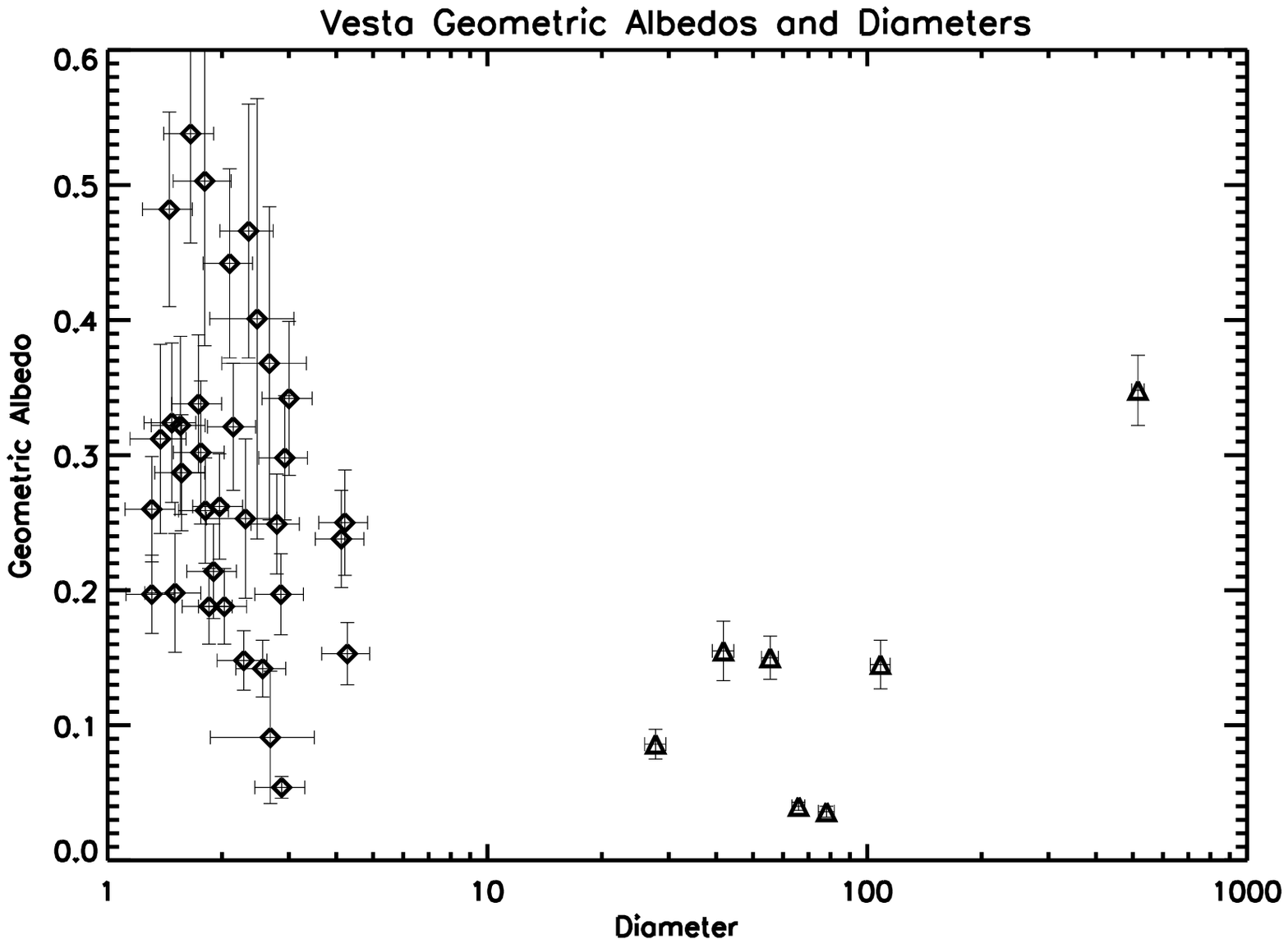}
\caption{Left: Flora family albedo distribution as a function of diameter. Flora family asteroids with diameters $>$ 10 km are from \textit{IRAS}, those with diameters $<$ 10 km are from \textit{Spitzer}. Right: Vesta family albedo distribution as a function of diameter. Vesta family asteroids with diameters $>$ 20 km are from \textit{IRAS}, those with diameters $<$ 20 km are from \textit{Spitzer}. }
\label{fig:flora_fam}
\end{figure}

\begin{figure}
\epsscale{1.15}
\plottwo{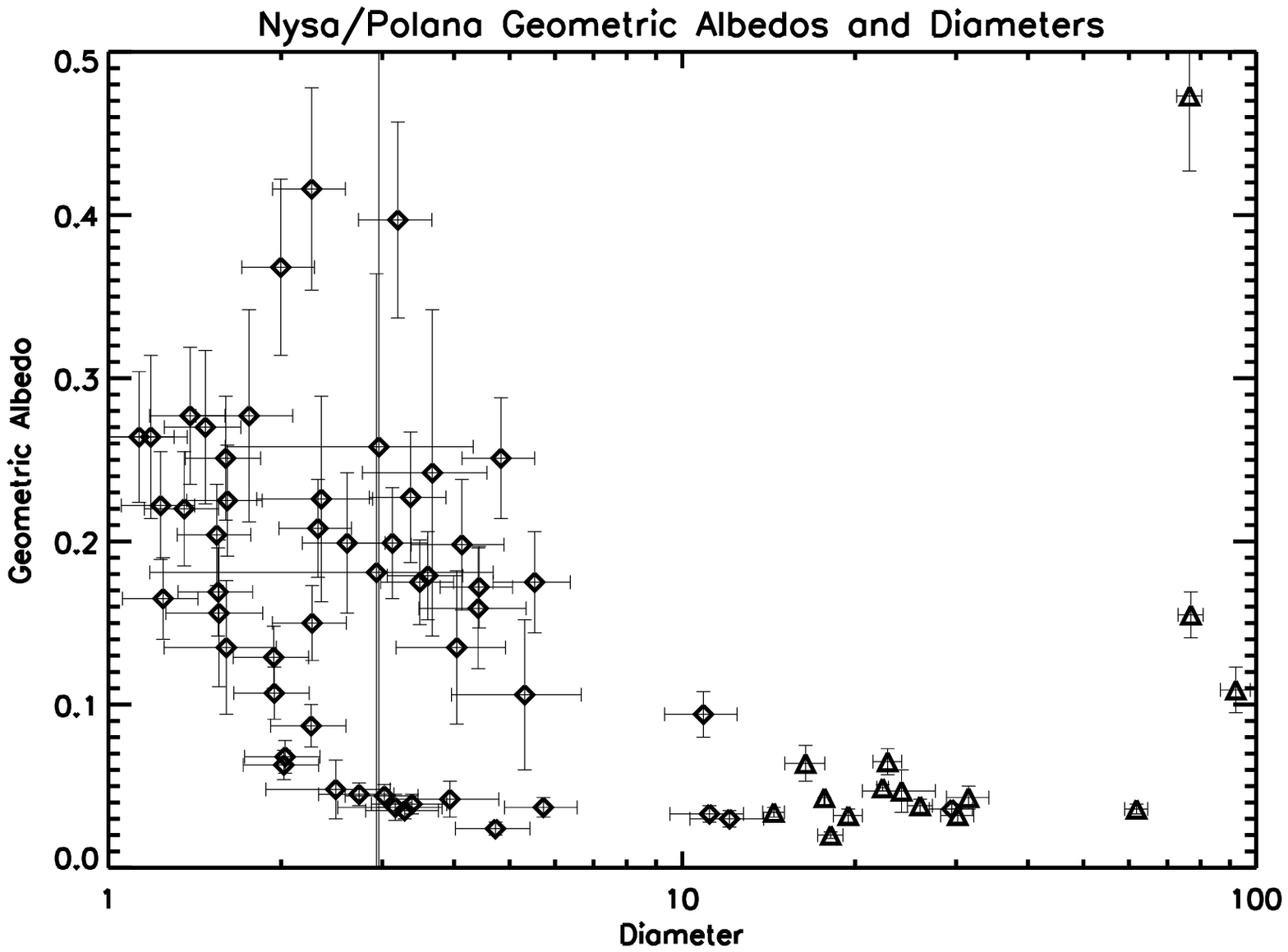}{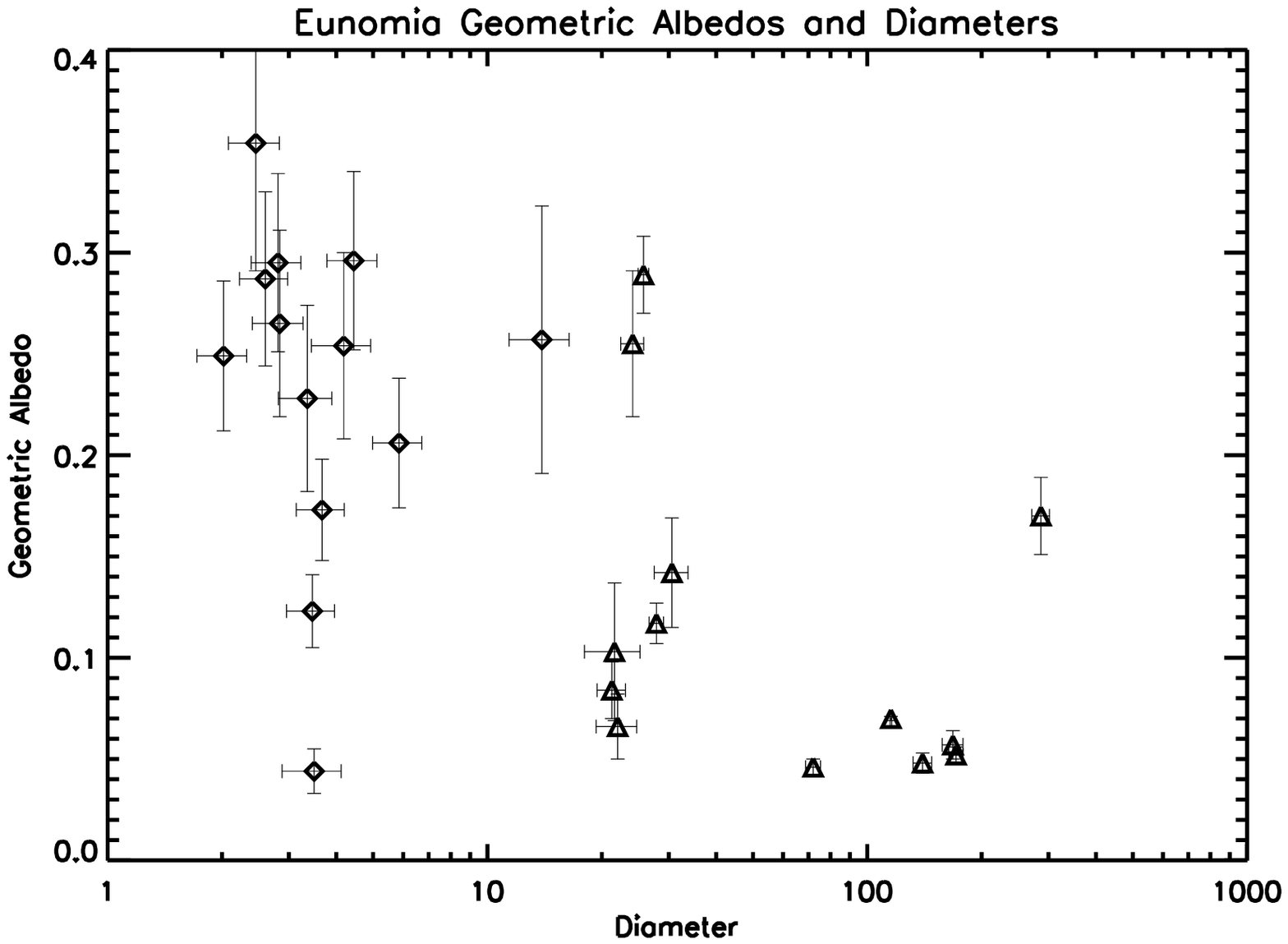}
\caption{Left: Nysa/Polana family albedo distribution as a function of diameter. Nysa/Polana family asteroids with diameters $>$ 10 km are from \textit{IRAS}, those with diameters $<$ 10 km are from \textit{Spitzer}. Right: Eunomia family albedo distribution as a function of diameter. Eunomia family asteroids with diameters $>$ 20 km are from \textit{IRAS}, those with diameters $<$ 20 km are from \textit{Spitzer}.}
\label{fig:nysa_fam}
\end{figure}

\begin{figure}
\epsscale{1.15}
\plottwo{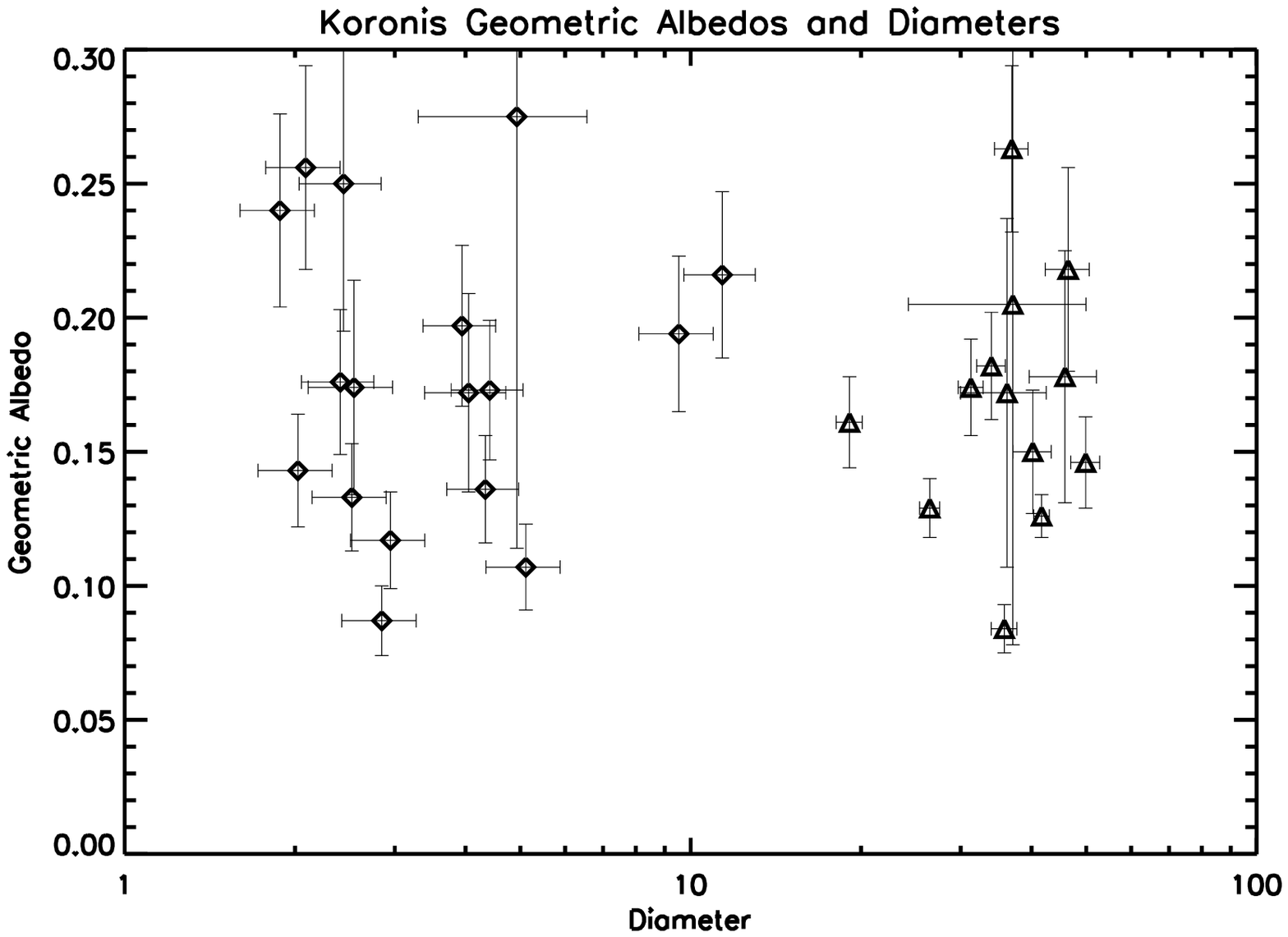}{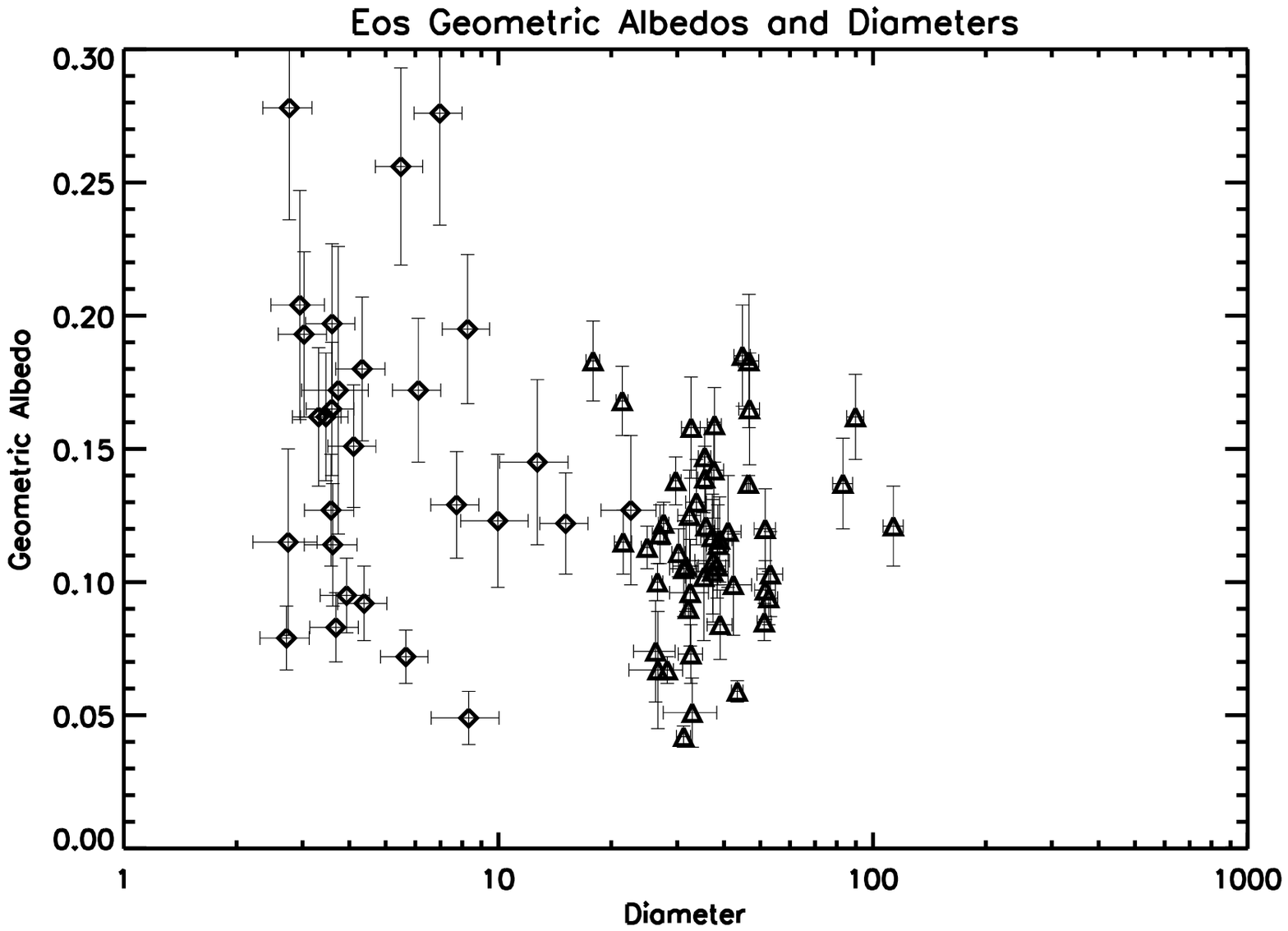}
\caption{Left: Koronis family albedo distribution as a function of diameter. Koronis family asteroids with diameters $>$ 20 km are from \textit{IRAS}, those with diameters $<$ 20 km are from \textit{Spitzer}. Right: Eos family albedo distribution as a function of diameter. Eos family asteroids with diameters $>$ 18 km are from \textit{IRAS}, those with diameters $<$ 18 km are from \textit{Spitzer}. }
\label{fig:koronis_fam}
\end{figure}

\begin{figure}
\epsscale{1.15}
\plottwo{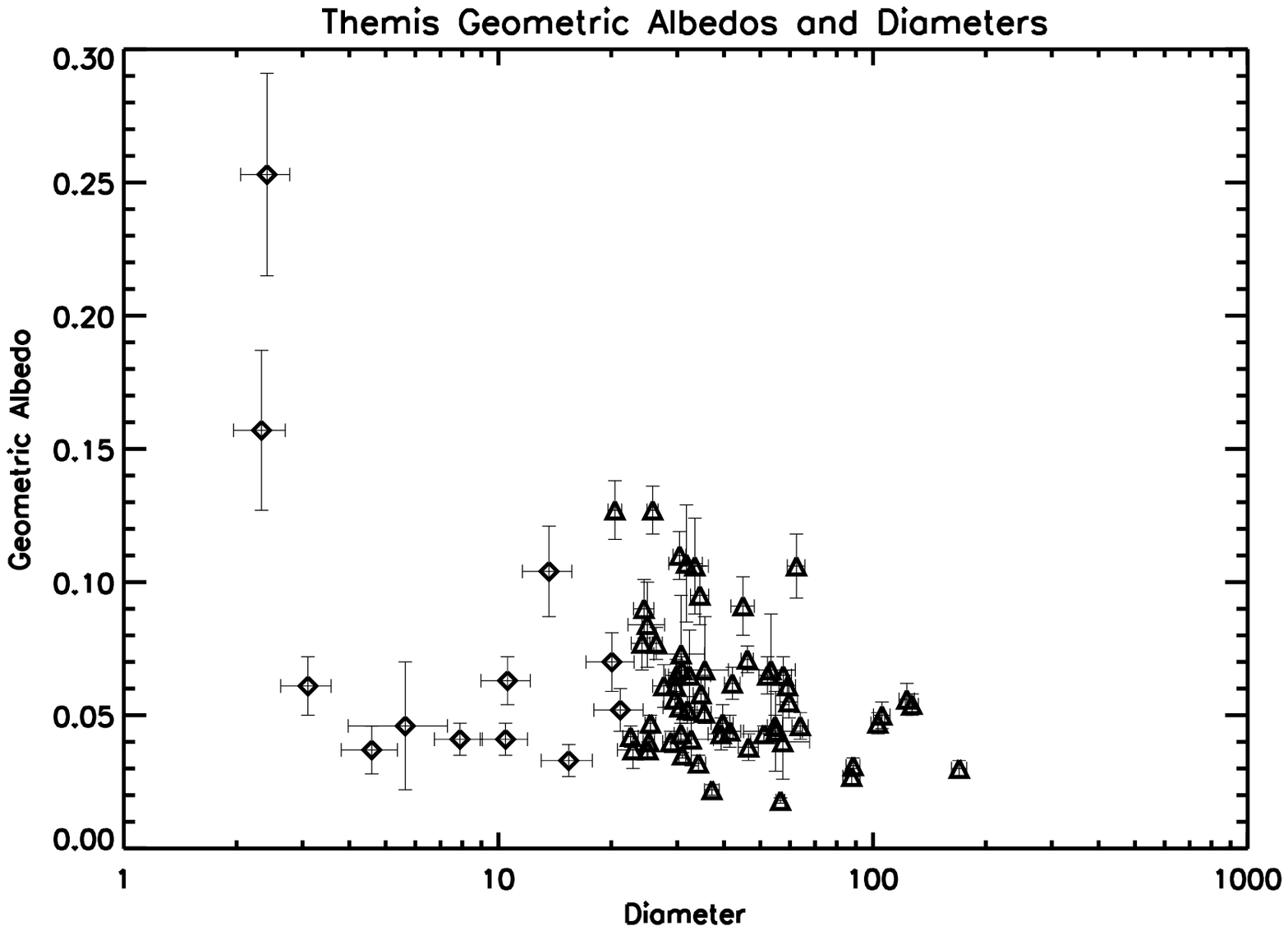}{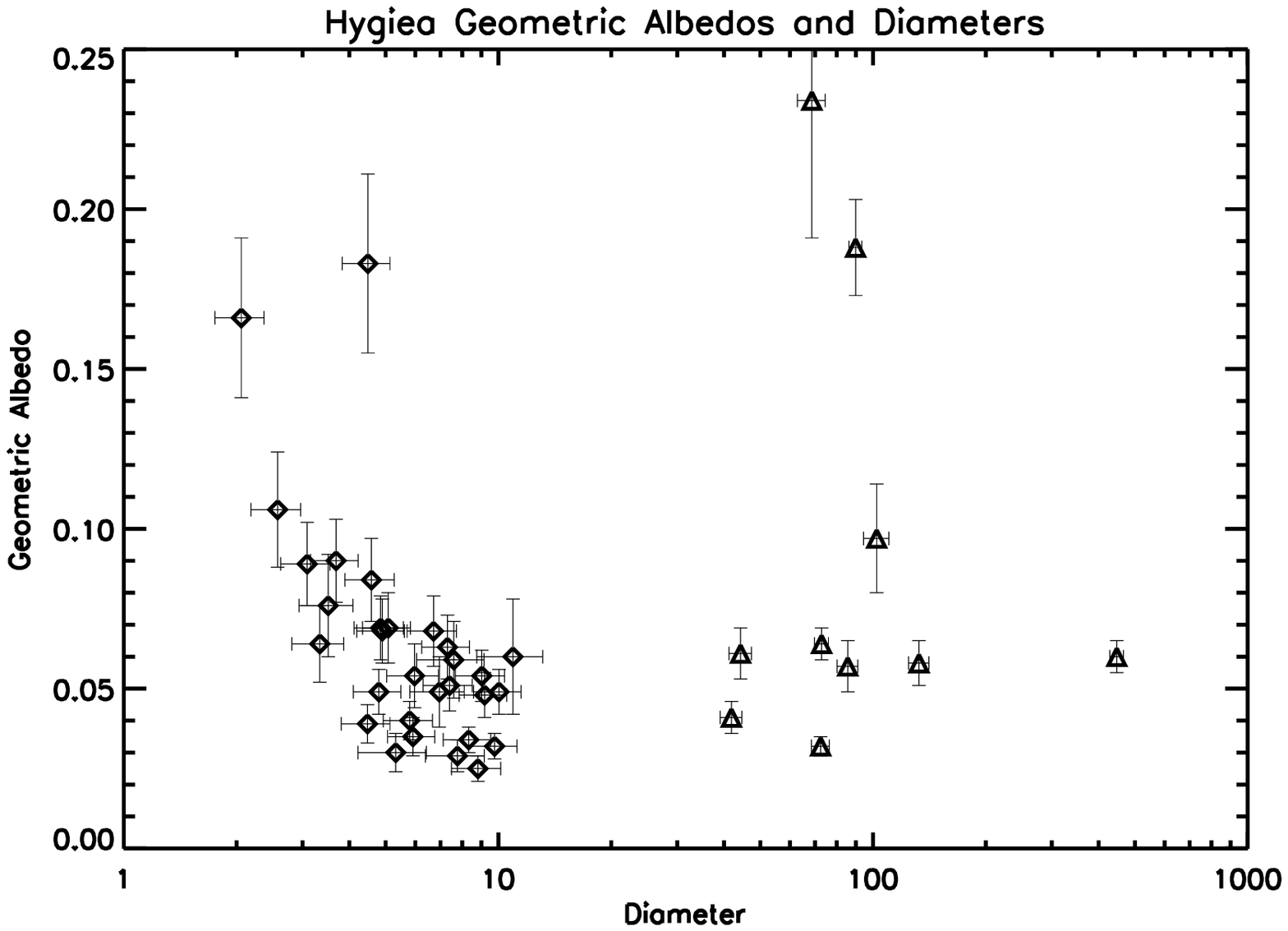}
\caption{Left: Themis family albedo distribution as a function of diameter. Themis family asteroids with diameters $>$ 20 km are from \textit{IRAS}, those with diameters $<$ 20 km are from \textit{Spitzer}. Right: Hygiea family albedo distribution as a function of diameter. Hygiea family asteroids with diameters $>$ 30 km are from \textit{IRAS}, those with diameters $<$ 10 km are from \textit{Spitzer}.}
\label{fig:themis_fam}
\end{figure}

Of the 47 dynamical families represented in the \textit{Spitzer} MIPSGAL 
and Taurus albedo catalog, eight Main Belt families have more than 20 
family members when combined with the IRAS and MSX albedo catalog. When 
albedo and diameter are compared for each dynamical family, no trends of 
increasing albedo with decreasing diameter are seen within the Main 
Belt population (Figures~\ref{fig:flora_fam} - \ref{fig:themis_fam}) and 
the mean albedos of the families are consistent with the taxonomic type 
of their largest member, except in the case of the Nysa/Polana family (Table ~\ref{table:ave_fam_alb}).
The Nysa/Polana family (Figure~\ref{fig:nysa_fam}) shows albedo 
evidence for what may be two taxonomic types within the family -- a very 
low albedo C-type asteroid grouping and a high albedo, S-type group. This 
split between the compositions of the family has been detected in the 
optical, where spectroscopic results found Nysa to be an S-type asteroid 
and Polana to be a C-type asteroid \citep{nysafam01}. Although 
spectroscopically it was unclear if this subdivision in compositional 
types extended to small diameters, we find evidence of both taxonomic 
types amongst the small family members.

The lack of albedo trend with decreasing diameter within the main 
belt potentially vitiates the origin of the albedo offset between the 
\textit{IRAS} and \textit{Spitzer} datasets arising from space weathering 
due to solar wind implantation on asteroid surfaces.  Although there is 
no direct way to measure asteroid age, a correlation between collisional 
timescale and asteroid diameter can be derived wherein the smaller 
asteroids are presumed to be on average younger than their larger 
neighbors \citep{davis2002}. To explain the 
color offsets between ordinary chondrites and 
S-type asteroids, lunar-type space weathering from the solar wind 
irradiation has been preferred mechanism invoked 
\citep[e.g.,][]{sunshine2004, Nesvorny2005, marchi2006} to account for 
the reddening of S-type asteroid slopes and a decrease in albedo 
with increasing asteroid age. The effects of space weathering are 
likely best understood by examining an asteroid population with a 
presumed common origin, such as the Koronis dynamical family,
whose \textit{Spitzer} derived albedos are presented in 
Figure~\ref{fig:koronis_fam}, combined with the optical colors of 
asteroids within the Koronis dynamical family \citep{Thomas2010}, which 
indicate a trend towards a redder optical slope with increasing 
diameter. No trend towards an increased albedo is apparent in the 2 to 5~km 
diameter Koronis family population observed by \textit{Spitzer}, 
although \citet{Thomas2010} argue that of a trend towards bluer colors 
exists in this size range. This trend, where the color reddens as a 
function of age but the geometric albedo shows little to no 
modification, is commensurate with the interpretation \textit{Galileo} 
flyby data of Koronis family member, 243~Ida \citep{chapman96, helfenstein96}.  
The \textit{Galileo} imaging data of 243~Ida shows large variations in 
spectroscopic absorption band depths and 1 and 2~\micron{} related to 
$^{+2}$Fe; however, there is a lack of albedo variation commensurate with 
the varying surface color. Individual $\simeq 52$~nm diameter, 
irregularly shaped dust particle samples returned from the surface of the 
S-type NEA Itokawa have amorphous rims populated by \textit{small} 
nanophase iron particles with an average size of $\sim 2$~nm 
\citep{noguchi2011}. Laboratory measurements of small ($< 10$~nm) nanophase 
iron particles indicate that these particles only redden reflectance 
spectra, and their presence in asteroid regolith would not result in a 
decreased albedo \citep{keller1999}. To produce a reduction (darkening) 
of the albedo and a steeper slope to the reflectance spectra (reddening) 
requires vapor deposition of ``larger'' ($\ltsimeq 40$~nm) metallic 
iron nanoparticles on grain rims \citep{noble2007, pieters2000}. 
The variation of spectroscopic band depths related to $^{+2}$Fe on 
243 Ida without a related color variation could also be a 
signature derived from small nanophase iron particles deposited on 
the surfaces of individual regolith dust particles. In our 
\textit{Spitzer} dataset, no trend is evident correlating an increasing 
albedo with decreasing diameter for the Koronis family and the other 
S-type families, including Flora, Eunomia and Eos. Hence, invoking 
traditional lunar-type space weathering mechanisms alone may not be 
sufficient to explain the relatively large albedo diversity within the 
small main belt asteroid population.

Our results wherein albedo does not change with diameter, and therefore 
age, coupled with results from \textit{Galileo} and the 
\textit{Hyabusa} mission suggests that the dominant space weathering 
mechanism is one which produces small nanophase iron particles. 
Lunar-type space weathering cannot be directly ruled out as the 
mechanism which causes space weathering within the asteroid 
belt. However as small nanophase iron particles do not modify 
the albedo, space weathering which produces these particles is 
insufficient to explain the relatively large fraction of small 
MBAs in the high albedo ($p_{V} > 0.15$) tail of the 
main belt asteroid albedo distribution. This observation suggests 
that the high albedo tail of the MBA albedo distribution is a 
function of composition, rather than space weathering.

\subsection{Size-Frequency Distributions}

\begin{figure}
\epsscale{0.8}
\plotone{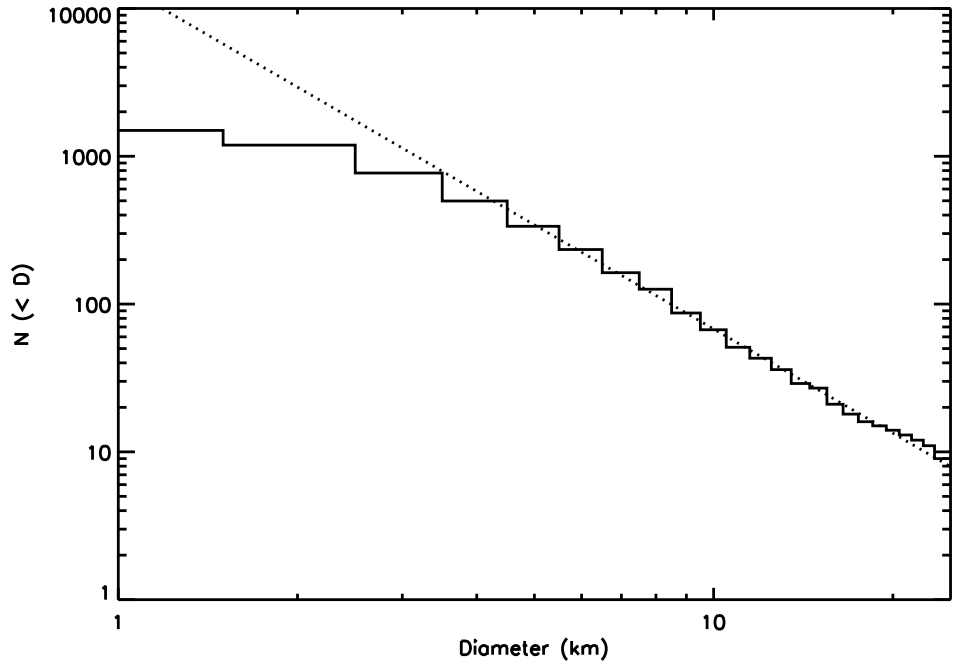}
\caption{The cumulative size-frequency distribution (SFD) of the \textit{Spitzer} catalogs is represented by the solid line. The dotted line represents the small diameter power-law fit where $b_{1}= 2.02 \pm 0.06$, the dashed line represents the large diameter power-law fit where $ b_{2}= 3.04 \pm0.05$.}
\label{fig:cumulative_sfd}
\end{figure}

From the MIPSGAL and Taurus data we can directly derive a 
size-frequency distribution (SFD) slope for small asteroids. Optical 
surveys such as the SDSS \citep{ivezic01} and Spacewatch \citep{jedicke98} 
have derived size-frequency distributions of main belt asteroids by 
assuming a mean albedo for all observed objects.  The slope of the 
cumulative size frequency distributions, $b$, from the relation 
$N ( > D) \propto D^{-b}$, as derived by these two surveys ranges 
from $b= 1.3$ to 1.8 respectively over an optical magnitude range 
$V \le 21$. The cumulative SFD for all main belt asteroids in the 
\textit{Spitzer} MIPSGAL and Taurus surveys is presented in 
Figure~\ref{fig:cumulative_sfd}.  The SFD between 7 and 25 km can be fit by a single power-law slope of b=2.34 $\pm$0.05. The measured SFD deviates from this fitted slope by 3$\sigma$ starting at 8 km. If one assumes that the break is a result of optical survey completeness, rather than a signature of the a transition between the regimes of  asteroid strength dominated by material strength versus gravitational potential energy as predicted from laboratory studies \citep{holsapple94,housen90}, then it can be said that current optical asteroid surveys are only complete to $\sim$ 8 km. Removal of dynamical families from the \textit{Spitzer} dataset modifies the power-law slope slightly; however, these changes in $b$ are less than the 
derived uncertainties ($\pm 0.03$). 

\begin{figure}
\epsscale{0.8}
\plotone{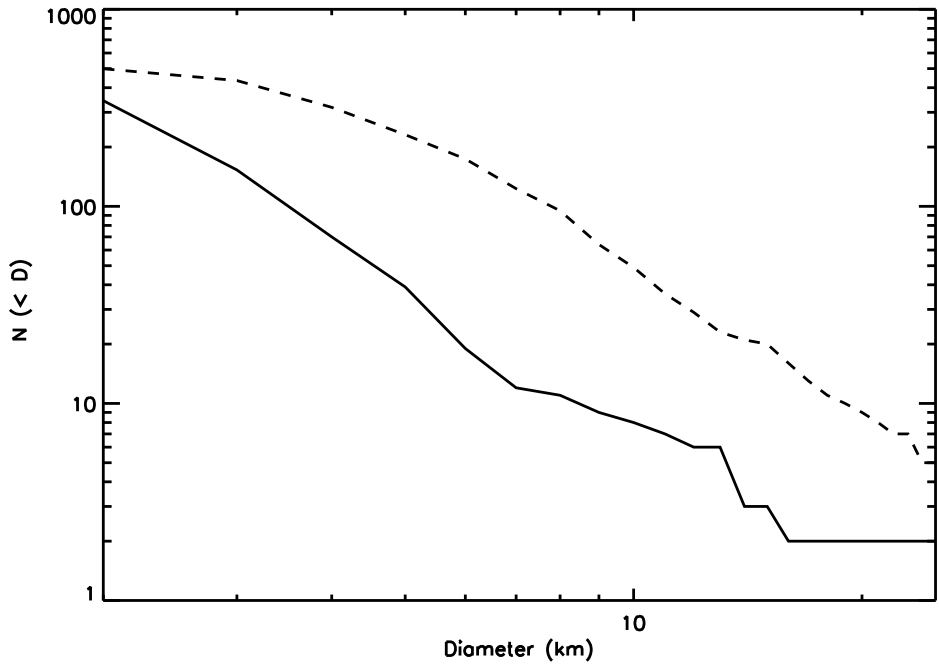}
\caption{The cumulative size-frequency distribution (SFD) of S-type (solid line) and C-type (dashed line) asteroids from the \textit{Spitzer} MIPSGAL and Taurus survey asteroid catalog.}
\label{fig:s_v_c_type_sfd}
\end{figure}

A difference between power-law SFD slopes of asteroids was noted 
in the g$^{\prime}$ and r$^{\prime}$ filter surveys by \citet{wiegert07}. 
Although it was unclear if this was an effect of color or albedo, 
the \citet{wiegert07} result can be tested with the MIPSGAL and Taurus 
data by using albedo as a proxy for composition. We have utilized the 
albedo ranges from \citet{ryan2010} for S- and C-type taxonomic groups and 
present the SFDs in Figure~\ref{fig:s_v_c_type_sfd}. The slope of the 
C-type SFD between 8 and 25~km is $ b =  2.49~\pm~0.07$, far shallower 
than the SFD slope of $b = 2.20~\pm~0.18$ derived for the S-type asteroids 
between 8 and 25~km in the MIPSGAL and Taurus catalogs. These 
\textit{Spitzer} results are similar to the slopes derived 
by \citet{wiegert07}, indicating that difference between the SFD slopes 
derived in g$^{\prime}$ and r$^{\prime}$ filters were likely a function 
of composition/taxonomic type.

\section{CONCLUSIONS}\label{sec:conclusions}

From the study of small Main Belt asteroids with \textit{Spitzer}, we 
find that some these objects are more diverse than the large main belt 
asteroids observed by \textit{IRAS} and \textit{MSX}. The mean geometric 
albedo for small main belt asteroids is higher than that of 
large main belt asteroids and the overall range of albedo variation is 
greater for small asteroids by a factor of 2. The distribution of low 
albedo asteroids in the solar system is also very different for small 
and large asteroids; only 9\% of all large (D $>$ 10~km) asteroids with 
C-type albedos are found in the inner Main Belt, but 24\% of all 
small ($D < 10$~km) asteroids with C-type albedos are found in the 
inner main belt. 

Though the extreme diversity of main belt asteroid albedos could be 
attributed to space weathering effects, this interpretation is not 
supported by the albedo results within dynamical families. Of the eight 
main belt dynamical families with more than 20 objects in the 
\textit{Spitzer} and IRAS catalogs, none show the clear relationship 
of increasing albedo with decreasing diameter characteristic of 
lunar-type space weathering. To determine if this diverse albedo range 
is caused by space weathering or compositional variations optical 
colors and/or spectra of these small main belt asteroids will be 
required to discriminate compositional taxonomies.

The bulk size-frequency distribution (SFD) of the Main Belt utilizing asteroid 
diameters was derived directly from the \textit{Spitzer} survey data. This 
bulk SFD shows evidence for a power-law break at 8.62~km. This asteroid 
diameter is consistent with the break diameter found for the Hilda group 
asteroid population \citep{ryan2010} and suggests that asteroid diameters 
of $\simeq 8.5$~km lie at the transition boundary where smaller bodies 
are dominated by internal material strength, whereas larger bodies are 
bound by gravitational potential energy. This SFD break 
derived from measures of the small asteroid population (diameters down 
to $\sim 1$~km) occurs at larger diameters than those suggested from 
dynamical modeling of the evolution of these bodies \citep{bottke05}. 
Our \textit{Spitzer} results therefore provide new observational 
constraint for collisional models that purport to follow the 
evolution of rocky planetesimals over the lifetime of the solar system.

\acknowledgements
ELR and CEW acknowledge support from National Science Foundation 
grant AST-0706980 to conduct this research. This work is based, 
in part, on archival data obtained with the 
\textit{Spitzer Space Telescope}, which is operated by the Jet 
Propulsion Laboratory, California Institute of Technology under a 
contract with NASA. Support for this work was provided by an award 
issued by JPL/Caltech. The authors also wish to thank the meticulous reading of our manuscript by an anonymous referee whose insight helped to improve the narrative.

%
%
%
%
%
%


%
%
%
%


\begin{deluxetable}{lccccccccccccccr}
\rotate
\tabletypesize{\tiny}
\setlength{\tabcolsep}{1pt}
\tablewidth{0pt}

\tablecaption{Orbital elements and 24~\micron{} fluxes for asteroids detected in the MIPSGAL and Taurus Surveys
\label{table:flux_table}}
\tablehead{
\colhead{Name or} & \colhead{Request} & \colhead{Date} & \colhead{Time} & \colhead{Predicted} & \colhead{Predicted} & \colhead{Detected} & \colhead{Detected} & \colhead{Flux} & \colhead{24 \micron} & \colhead{Flux} & \colhead{Heliocentric} & \colhead{Geocentric} & \colhead{Phase} & \colhead{Absolute} & \colhead{Absolute}\\
\colhead{Provisional} & \colhead{Number} & \colhead{} & \colhead{} & \colhead{RA} & \colhead{Dec} & \colhead{RA} & \colhead{Dec} & \colhead{Flag} & \colhead{Flux} & \colhead{Uncertainty} & \colhead{Distance} & \colhead{Distance} & \colhead{Angle} & \colhead{Magnitude} & \colhead{Magnitude}\\
\colhead{Designation} & \colhead{} & \colhead{(UT)} & \colhead{(UT)} & \colhead{(Deg)} & \colhead{(Deg)} & \colhead{(Deg)} & \colhead{(Deg)} & \colhead{} & \colhead{(mJy)}  & \colhead{(mJy)} & \colhead{(AU)} & \colhead{(AU)}  & \colhead{(Deg)} &\colhead{} & \colhead{$\pm$}\\
}
\startdata
    1998KQ42&  15598848&  2005-09-28&  07:28:19.89& 273.74063& -15.64690& 273.75385& -15.65198 &   1&    24.950&    0.345&     1.72&	1.21&	12.03&	15.300&    0.150\\
     2000AF141&  15598848&  2005-09-28&  07:28:19.89& 273.74564& -17.95830& 273.75189& -17.95803&    1&    10.390&    0.337&     2.17&	1.73&	16.66&	15.100&    0.150\\
       1998QD70&  15598848&  2005-09-28&  07:28:19.89& 273.84332& -17.12060& 273.84808& -17.12097&    1&    22.830&    0.432&     2.81&	2.44&	20.02&	14.100&    0.150\\
       2001TT94&  15598848&  2005-09-28&  07:28:19.89& 273.79892& -15.39880& 273.80466& -15.39959&    1&    11.340&    0.315&     2.87&	2.50&	22.72&	15.000&    0.150\\
      2001TK102&  15598848&  2005-09-28&  07:28:19.89& 273.80942& -17.00150& 273.81311& -17.00447&    1&    20.590&    0.469&     3.47&	3.14&	16.88&	14.300&    0.150\\
       2006RG52&  15598848&  2005-09-28&  07:28:19.89& 273.74466& -18.40490& 273.74939& -18.40561&    1&     5.754&    0.338&     2.67&	2.29&	22.16&	16.900&    0.150\\
\enddata
\tablecomments{In column 9, the flux flags are the following: 1= asteroid flux which matches within 30\% of flux in subsequent epochs, 2= flux which varies $>$30\% btwn epochs, 3= nearby bright source, 4=blended bright source. The full catalog is will be available as a machine-readable table.}
\end{deluxetable}


\begin{deluxetable}{lccccr}
\tablecaption{Sighting solutions for asteroids detected in the MIPSGAL and Taurus Surveys
\label{table:sighting_solns}}
\tablehead{
\colhead{Name or} & \colhead{Request} & \colhead{Geometric} & \colhead{Geometric} & \colhead{Diameter} & \colhead{Diameter}\\
\colhead{Provisional} & \colhead{Number} & \colhead{Albedo} & \colhead{Albedo} & \colhead{} & \colhead{Error}\\
\colhead{Designation} &\colhead{} & \colhead{} & \colhead{Error} & \colhead{(km)} & \colhead{(km)} \\
}
\startdata
       1998KQ42&  15598848&     0.35&	0.04&	  1.95&     0.20\\
      2000AF141&  15598848&     0.36&	0.04&	  2.12&     0.22\\
       1998QD70&  15598848&     0.15&	0.02&	  5.23&     0.53\\
       2001TT94&  15598848&     0.12&	0.01&	  3.86&     0.39\\
      2001TK102&  15598848&     0.06&	0.01&	  7.38&     0.75\\
       2006RG52&  15598848&     0.06&	0.01&	  2.35&     0.24\\
\enddata
\tablecomments{The full catalog is will be available as a machine-readable table.}
\end{deluxetable}


\begin{deluxetable}{lccccc}
\tablecaption{Mean asteroid geometric albedos and diameters
\label{table:mean_solns}}
\tablehead{
\colhead{Name or} & \colhead{Geometric} & \colhead{Geometric} & \colhead{Diameter} & \colhead{Diameter} &\colhead{Number}\\
\colhead{Provisional} & \colhead{Albedo} & \colhead{Albedo} & \colhead{} & \colhead{Error} & \colhead{Observations}\\
\colhead{Designation} & \colhead{} & \colhead{Error} & \colhead{(km)} & \colhead{(km)} & \colhead{} \\
}
\startdata

  	1321T-2&    0.20&    0.05&	1.71&	  0.34&       2\\
  	1413T-2&     0.04&     0.01&	1.74&	  0.26&       1\\
       1978VE10&     0.08&     0.01& 	8.39&	  1.26&       1\\
  	1978VE6&     0.14&     0.04& 	2.96&	  0.52&       2\\
  	1978VZ5&     0.18&     0.03&	 4.79&	  0.70&       2\\
  	1979MM2&     0.39&     0.06&	1.69&	  0.25&       1\\
  	1980FY2&     0.212&     0.03&	 5.46&	  0.82&       1\\
       1981EA29&     0.05&     0.01&	 8.29&	  1.26&       2\\
\enddata
\tablecomments{The full catalog is will be available as a machine-readable table.}
\end{deluxetable}

\begin{deluxetable}{lccccl}
\tabletypesize{\small}
\tablecaption{ Comparison of Spitzer derived albedos and diameters to IRAS/MSX and Occultation Diameters
\label{table:big_asteroids}}
\tablehead{
\colhead{Asteroid} & \colhead{MIPS} & \colhead{MIPS} & \colhead{IRAS/MSX} & \colhead{IRAS/MSX} &\colhead{Occultation}\\
\colhead{Name} & \colhead{Geometric} & \colhead{Diameter}  & \colhead{Geometric} & \colhead{Diameter} & \colhead{Diameter}\\
\colhead{} & \colhead{Albedo} & \colhead{(km)} & \colhead{Albedo} & \colhead{(km)} & \colhead{(km)}\\
}
\startdata

103 Hera & 0.20 $\pm$ 0.04 & 88.30 $\pm$ 8.51 & 0.19 $\pm$ 0.02 & 91.58 $\pm$ 4.14 & 89.1 $\pm$ 1.1\\
106 Dione & 0.07 $\pm$ 0.01 & 168.72 $\pm$ 8.89 & 0.07 $\pm$0.01 & 169.92 $\pm$7.86 & 176.7 $\pm$ 0.4\\
206 Hersilia & 0.06 $\pm$ 0.02 & 97.99 $\pm$ 7.40 & 0.06 $\pm$ 0.01 & 101.72 $\pm$ 5.18 & \\
233 Asterope & 0.10 $\pm$ 0.01 & 97.54 $\pm$ 10.32 & 0.08 $\pm$ 0.01 & 109.56 $\pm$ 5.04 & \\
283 Emma & 0.03 $\pm$ 0.01 & 145.44 $\pm$ 7.72 & 0.03 $\pm$0.01 & 145.70 $\pm$ 5.89 & 148.00 $\pm$16.26\\
318 Magdalena & 0.03 $\pm$ 0.01 & 105.32 $\pm$ 11.11 & 0.03 $\pm$ 0.01 & 106.08 $\pm$ 0.25 &\\
1064 Aethusa & 0.17 $\pm$0.04 & 25.42 $\pm$4.28 & 0.27 $\pm$0.03 & 20.64 $\pm$1.37&\\
1122 Neith & 0.34 $\pm$ 0.02 & 13.81 $\pm$0.73 & 0.34 $\pm$ 0.07 & 13.84 $\pm$ 1.46 &\\
2007 McCuskey & 0.03 $\pm$0.01 & 35.26 $\pm$3.74 & 0.07 $\pm$ 0.01 & 33.79 $\pm$1.31&\\

\enddata
\end{deluxetable}


\begin{deluxetable}{cccc}
\tablecaption{Completeness limits in the optical and 24 \micron
\label{table:completeness_table}}

\tablehead{
 \colhead{Semimajor} &
 \colhead{Optical} &
 \colhead{Optical} &
  \colhead{24 \micron}\\
   
   \colhead{Axis} & 
   \colhead{Completeness} & 
   \colhead{Completeness} & 
   \colhead{Completeness}\\

 \colhead{Range} & 
 \colhead{$H$} & 
 \colhead{Diameter} & 
 \colhead {Diameter}\\
 
\colhead{(AU)} & 
\colhead{(mag)} & 
\colhead{(km)} & 
\colhead{(km)}\\
}
\startdata
2.06 - 2.5 & 17.25 & 3.33 & 0.79\\
2.5 - 2.82 & 16.75 & 4.20  & 1.05\\
2.82- 3.27 & 16.25 & 5.28 & 1.47\\
3.27 - 3.65 & 15.75 & 6.65 & 1.88\\
\enddata
\end{deluxetable}


\begin{deluxetable}{lccccr}
\tablecaption{Geometric Albedos of Dynamical Families Derived from \textit{Spitzer} Surveys
\label{table:ave_fam_alb}}

\tablehead{
\colhead{Dynamical}&
\colhead{Heliocentric}&
\colhead{DeMeo}&
\colhead{Tholen}&
\colhead{Number of} &
\colhead{Mean Geometric}
\\
\colhead{Family}&
\colhead{Distance}&
\colhead{Taxonomic} &
\colhead{Taxonomic} &
\colhead{Members} &
\colhead{Albedo}\\

\colhead{}&
\colhead{(AU)}&
\colhead{Type} &
\colhead{Type} &
\colhead{} &
\colhead{}\\

}

\startdata
Flora & 2.20& Sw & S & 47 & 0.207 $\pm$ 0.092\\
Vesta & 2.36 & V & V & 42 & 0.272 $\pm$ 0.156\\
Nysa/Polana & 2.42 & S/B &  & 68 & 0.146 $\pm$ 0.107\\
Eunomia & 2.64 & K & S & 26 & 0.175 $\pm$ 0.101\\
Koronis & 2.87 & S & S & 30 & 0.174 $\pm$ 0.051\\
Eos & 3.01 & K & S & 78 & 0.128 $\pm$ 0.047\\
Themis & 3.13 & C & C & 71 & 0.063 $\pm$ 0.035\\
Hygiea & 3.14 & C & C&  40 & 0.071 $\pm$ 0.046\\
\enddata
\end{deluxetable}

\clearpage
%
%
%
%


\clearpage


\clearpage


\clearpage

\clearpage


\clearpage

\clearpage

\clearpage

 
\clearpage


 
\clearpage


\end{document}